\documentclass[aps,prx,twocolumn,nofootinbib,
amsmath,amssymb,amsfonts,eqsecnum,superscriptaddress]{revtex4}
\usepackage{amsmath}
\usepackage{bbm}
\usepackage{amssymb}
\usepackage{amsmath}
\usepackage{ifpdf}
\usepackage{color}
\usepackage{graphicx,epsfig}
\usepackage{hyperref}
\allowdisplaybreaks

\graphicspath{
{./}
{./FIGS/}
{../FIGS/}
}

\begin{document}
\title{Minimal model for Hilbert space fragmentation with local constraints}
\author{Bhaskar Mukherjee}
\affiliation{International Centre for Theoretical Sciences, Tata Institute of Fundamental Research, Bengaluru 560089, India}
\affiliation{Wilczek Quantum Center, School of Physics and Astronomy, Shanghai Jiao Tong University, Shanghai 200240, China}
\affiliation{Department of Physics and Astronomy, University of Pittsburgh, Pittsburgh, PA 15260, USA}
\author{Debasish Banerjee}
\affiliation{Saha Institute of Nuclear Physics, HBNI, 1/AF Bidhannagar, Kolkata 700064, India}
\author{K. Sengupta}
\affiliation{School of Physical Sciences, Indian Association for the Cultivation of Science, Kolkata 700032, India}
\author{Arnab Sen}
\affiliation{School of Physical Sciences, Indian Association for the Cultivation of Science, Kolkata 700032, India}

\begin{abstract}

Motivated by previous works on a Floquet version of the PXP model
[Mukherjee {\it et al.} Phys. Rev. B 102, 075123 (2020), Mukherjee
{\it et al.} Phys. Rev. B 101, 245107 (2020)], we study a
one-dimensional spin-$1/2$ lattice model with three-spin
interactions in the same constrained Hilbert space (where all
configurations with two adjacent $S^z=\uparrow$ spins are excluded).
We show that this model possesses an extensive fragmentation of the
Hilbert space which leads to a breakdown of thermalization
upon unitary evolution starting from a large class of simple
initial states. Despite the non-integrable nature of the
Hamiltonian, many of its high-energy eigenstates admit a
quasiparticle description. A class of these, which we dub as
``bubble eigenstates'', have integer eigenvalues (including
mid-spectrum zero modes) and strictly localized quasiparticles while
another class contains mobile quasiparticles leading to a dispersion
in momentum space. Other anomalous eigenstates that arise due to a
{\it secondary} fragmentation mechanism, including those that lead
to flat bands in momentum space due to destructive quantum
interference, are also discussed. The consequences of adding a
(non-commuting) staggered magnetic field and a PXP term respectively
to this model, where the former preserves the Hilbert space
fragmentation while the latter destroys it, are discussed.
Making the staggered magnetic field a periodic function of time
  defines an interacting Floquet system that also evades thermalization and
  has additional features like exact stroboscopic freezing of an
exponentially large number of initial states at special drive frequencies.
Finally, we map the model to a $U(1)$ lattice gauge theory coupled
to dynamical fermions and discuss the interpretation of some of
these anomalous states in this language. A class of gauge-invariant
states show reduced mobility of the elementary charged excitations
with only certain charge-neutral objects being mobile suggesting a
connection to fractons.
\end{abstract}
\maketitle

\section{Introduction}

Given a local Hamiltonian, the eigenstate thermalization hypothesis
(ETH) posits that the reduced density matrix for a subsystem
corresponding to any finite energy density eigenstate in a
thermodynamically large quantum system is {\it thermal}~\cite{ETH1,
ETH2, ETH3}. ETH also explains why such a system locally (but not
globally) reaches thermal equilibrium under its own unitary
dynamics~\cite{ETH3, ETH4, ETH5, ETH6, ETH7} when initially prepared
in a far-from-equilibrium pure state (which, therefore, stays pure
at all times) with the corresponding temperature determined by its
energy density. It is equally interesting to explore how ETH might
be violated in interacting quantum many-body systems.
Integrability~\cite{VR2016} and many-body localization~\cite{PH2010,
NH2015} (where the former requires fine-tuning of interactions while
the latter requires strong disorder) provide two well-known routes
where the presence of an extensive number of emergent conserved
quantities (in both cases) causes a failure to
thermalize and retention of memory starting from all initial
conditions.

Can non-integrable disorder-free systems also show an intermediate
dynamical behavior between thermalization in generic interacting
systems and its complete breakdown for integrable/many-body
localized ones? A weaker form of ergodicity breaking was indeed
experimentally observed in a one-dimensional quantum simulator
composed of $51$ Rydberg atoms~\cite{ryd_exp}. In
such one-dimensional (1D) Rydberg chains, while most initial states
rapidly thermalized in accordance with ETH, a high-energy N\'eel
state of Rydberg atoms alternating between their ground and excited
state respectively on the lattice (denoted by $|\mathbb{Z}_2\rangle$
henceforth) instead showed persistent oscillations in spite of the
large Hilbert space dimensionality of the system. Subsequent
theoretical works~\cite{pxp1,pxp2} used an effective spin-$1/2$
model (where $S_i^z=\uparrow (\downarrow)$ denotes a Rydberg excited
state (ground state) for the atom at site $i$), the so-called PXP
model~\cite{pxp3, pxp3a, pxp4}, in a constrained Hilbert space where
no two adjacent sites in a chain can have $S^z=\uparrow$ together to
mimic the strong Rydberg blockade present in the experiment. The
spectrum of the PXP chain contains an extensive number of highly
athermal high-energy eigenstates~\cite{pxp1,pxp2,pxp5}, called
quantum many-body scars, embedded in an otherwise thermal spectrum
of ETH-satisfying eigenstates. These quantum scars are almost
equally spaced in energy and are responsible for the oscillations
starting from a $|\mathbb{Z}_2\rangle$ state~\cite{pxp1,pxp2,pxp5}
due to their high overlap with it. Apart from the PXP model, a large
variety of other systems have now been shown to exhibit quantum
many-body scarring, including the AKLT chain~\cite{aklt1, aklt2,
aklt3,aklt4,aklt5}, quantum Hall systems in the thin torus
limit~\cite{qh1,qh2}, fermionic Hubbard
model~\cite{HM1,HM2,HM3,HM4}, spin $S=1$ magnets~\cite{mag1}, driven
quantum matter~\cite{dyn1, dyn2, dyn3, dyn4, dyn5, dyn6, dyn7},
two-dimensional Rydberg systems~\cite{2d_ryd1, 2d_ryd2},
geometrically frustrated magnets~\cite{geometric1,geometric2} and
certain lattice gauge theories~\cite{LGTscars} (for a recent review,
see Ref.~\onlinecite{scarsreview}).

Another ETH-violating mechanism may arise in models where the
presence of certain dynamical constraints results in a restricted
mobility of the excitations~\cite{Chamon2005, Haah2011, VijayHF2016,
Pretko2017}. It was shown in Refs.~\onlinecite{PaiPN2019,
KhemaniN2019, Khemani2020, SalaRVKP2020, MoudgalyaPNRB2019} that imposing both
``charge'' and ``dipole moment'' conservations lead to an emergent
fragmentation of the Hilbert space into exponentially many
disconnected subspaces. Crucially, even states within the same
symmetry sector (i.e., with same values of total charge and dipole
moment respectively) may become dynamically disconnected whereby the
unitary evolution only connects states within smaller
``fragments''. Each of these fragments only occupies a
vanishing fraction of the total Hilbert space.
Ref.~\onlinecite{YangLGI2020} showed that fragmentation could
also be achieved in 1D spin $S=1/2$ systems due to a strict
confinement of Ising domain walls (see also
Ref.~\onlinecite{Tomasi2019, LanglettX2021}).
Recently, another work considered
a spin $S=1$ version of the PXP model in a suitably defined
constrained Hilbert space~\cite{MukherjeeCL2020} to show that
fragmentation can also arise in such settings. Finally, a
quasi-one-dimensional geometrically frustrated spin $S=1/2$
Heisenberg model with local conserved quantities on rungs of the
ladder~\cite{geometric2} induced Hilbert space fragmentation into
sectors composed of singlets and triplets on
rungs~\cite{HahnCL2020} (see Ref.~\onlinecite{LeePC2020}
  for another frustration-induced Hilbert space fragmentation mechanism).
A recent experiment in an ultra-cold atomic
quantum simulator that realized a tilted 1D
Fermi-Hubbard model~\cite{expfragmentation} observed non-ergodic
behavior possibly due to Hilbert space fragmentation.

In this work, we will consider a 1D spin $S=1/2$ model with
three-spin interactions defined in the same constrained Hilbert
space as the original $S=1/2$ PXP model. This model arises as the
dominant non-PXP type interaction term in a perturbative expansion
of the Floquet Hamiltonian when a certain periodically driven
version of the PXP model is considered~\cite{dyn1,dyn5}. Like the
PXP model, this model has a spectrum which is symmetric around zero
energy and has an exponentially large number of exact zero modes (in
system size) due to an index theorem linked to an intertwining of
particle-hole and inversion symmetry~\cite{indexth}. However, unlike
the PXP model, this model shows Hilbert space fragmentation due to a
combination of the kinematic constraints and the nature of the
interactions. The fragmentation results in a large number of exact
eigenstates with non-zero integer energies (with a suitable choice
of Hamiltonian normalization) apart from the zero modes.
Furthermore, the fragmentation allows us to write closed-form
expressions for many high-energy eigenstates for an arbitrary system
size since they admit an exact quasiparticle description. A class of
these, which we dub as ``bubble eigenstates'', have strictly
localized quasiparticles and integer eigenvalues (including zero).
The other exact ``non-bubble'' eigenstates cannot be written in
terms of localized quasiparticles and are more easily described in
momentum space. The simplest class of these have mobile
quasiparticles leading to an energy dispersion in momentum space.

We also show the presence of a novel {\it secondary}
fragmentation mechanism whereby certain linear combinations of a
{\it fixed number of basis states} in momentum space form smaller
emergent fragments in the Hilbert space. This secondary
fragmentation leads to perfectly flat bands of non-bubble
eigenstates in momentum space due to a destructive quantum
interference phenomenon, among other eigenstates. The signatures of
fragmentation also shows up in the unitary dynamics following a
global quench from simple initial states, with a class of states
showing perfect (undamped) oscillations of local correlations while
other initial conditions like the N\'eel state
($|\mathbb{Z}_2\rangle$) showing thermalization (instead of
scar-induced oscillations as in the original PXP model).

Next, we consider the effects of adding two
different non-commuting interactions to this model respectively, a
staggered magnetic field term and a PXP type interaction. The former
interaction preserves the Hilbert space fragmentation and remarkably
allows for many exact zero modes that are simultaneous eigenkets of
both the (non-commuting) terms in the Hamiltonian.
The latter
interaction destroys the Hilbert space fragmentation but with some
interesting differences between the cases where the PXP term can be
considered as a small perturbation and where it is not small
compared to the three-spin interaction.
We also consider a Floquet version with a time-periodic
  staggered magnetic field and show that the model {\it{does not}} locally
  heat up to an infinite temperature ensemble as expected of generic
  interacting many-body systems~\cite{FETH1,FETH2,FETH3}.
  The Floquet version has an
  emergent $SU(4)$ symmetry and furthermore, shows exact stroboscopic
  freezing of an
  exponentially large number of initial states at specific drive frequencies
  even in the thermodynamic limit.

The three-spin model in a constrained Hilbert space
that we study can be exactly mapped to a Hamiltonian formulation of
a $U(1)$ lattice gauge theory coupled to dynamical fermions.
Interestingly, only the total charge and not the net dipole moment is
conserved in this lattice gauge theory unlike previous
(fractonic) theories which show Hilbert space fragmentation due to
the conservation of higher moments. We give an interpretation of the
bubble eigenstates and some non-bubble eigenstates in this gauge
theory language. Furthermore, we demonstrate the reduced mobility of
certain charge-neutral units in these states suggesting an emergent
fracton-like picture.

The rest of the paper is arranged in the following manner. In
Section~\ref{sec:model}, we introduce the model and explain its
basic features. We then introduce the simplest class of exact
eigenstates of this model, the bubble eigenstates, which can be
written in terms of strictly localized excitations and discuss their
properties in Section~\ref{sec:bubblestates}. After that, we focus
on non-bubble exact eigenstates and discuss three types of such
states. In Section~\ref{sec:nonbubblestates}, we introduce the
simplest class of non-bubble eigenstates which can be understood
using bubbles that hybridize with the ``vacuum'' and become mobile.
In Section~\ref{sec:secondaryfrag}, we discuss another class of
excited eigenstates that arise due to a more subtle mechanism of
secondary fragmentation in the Hilbert space, with two examples
introduced in Section~\ref{subsec:k0I1} and
Sec~\ref{subsec:flatbands}. The latter class of states leads to the
formation of flat bands in momentum space, but in the high-energy
spectrum. We then discuss the consequences of adding further
non-commuting interaction terms to this model in
Section~\ref{sec:otherint} where we focus on two different cases. In
Section~\ref{subsec:staggered}, we consider the effects of adding a
staggered magnetic field that preserves the fragmentation of the
original model. We also discuss a class of exact zero modes for this
theory that are simultaneous eigenkets of both the non-commuting
interaction terms in the Hamiltonian.
In Section~\ref{subsec:Floquet}, we study a Floquet
  version of the same and show extra features like an
emergent $SU(4)$ symmetry for certain states and a drive-induced
exact dynamical freezing of an
exponentially large number of simple initial states at stroboscopic
times.
In Section~\ref{subsec:PXP},
we study the effects of adding a PXP term to the model and show that
such a term destroys the Hilbert space fragmentation of the original
model. Nonetheless, some signatures of fragmentation still survive
when the PXP term can be considered to be a small perturbation as we
will show. In Section~\ref{sec:quenches}, we will discuss global
quenches starting from some simple initial states (including the
$|\mathbb{Z}_2\rangle$ state) for the three-spin interaction model
and also when this model is supplemented by a staggered field term
and a PXP term, respectively. We show the mapping of the model to a
lattice gauge theory with $U(1)$ gauge fields and dynamical fermions
in Section~\ref{sec:LGT}. We show how to interpret a class of the
exact eigenstates with this gauge theory approach. We finally
conclude and highlight some open issues, including a possible
realization of this model using a Floquet setup, in
Section~\ref{sec:conclusions}.

\section{Model for Hilbert space fragmentation}
\label{sec:model} The 1D spin $S=1/2$ model that will serve as a
minimal model for Hilbert space fragmentation is as follows:
\begin{eqnarray}
H_3 &=& w_s \sum_{j=1}^L (\tilde{\sigma}_j^+ \tilde{\sigma}_{j-1}^- \tilde{\sigma}_{j+1}^- + \mathrm{H.c.}) \nonumber \\
&=& w_s \sum_{j=1}^L P^{\downarrow}_{j-2} \left( |\downarrow_{j-1}
\uparrow_{j} \downarrow_{j+1} \rangle \langle \uparrow_{j-1}
\downarrow_{j} \uparrow_{j+1}|\right. \nonumber\\
&& \left. +\mathrm{H.c.}\right) P^{\downarrow}_{j+2} \label{H3def}
\end{eqnarray}
where in the first line, $\sigma^\alpha_j$ represents Pauli
matrices at site $j$ for $\alpha=x,y,z$ and $\sigma_j^{\pm} =
(\sigma_j^x \pm i\sigma_j^y)/2$ with the projector $P^{\downarrow}_j
= (1-\sigma_j^z)/2$ and $\tilde{\sigma}_j^\alpha =
P^\downarrow_{j-1}\sigma_j^\alpha P^\downarrow_{j+1}$. In the second
line, $S_j^z=\uparrow_j, \downarrow_j$ represents the two states of
the spin on site $j$ (the computational basis). $L$ is taken to be
even and periodic boundary conditions assumed ($\sigma_{L+1}^\alpha
= \sigma_1^\alpha$). In the rest of this work, we shall set $w_s$ to
unity and scale all energies in units of $w_s$. This model is
defined in a constrained Hilbert space (the same one as that of the
paradigmatic PXP model) with the property that no two adjacent spins
can have $S^z=\uparrow$. This implies that the
number of Fock states ($|\cdots,S^z_j,S^z_{j+1},S^z_{j+2},
S^z_{j+3}, \cdots \rangle$) in the computational basis scales as
$F_{L-1}+F_{L+1}$ (where $F_n$ denotes a Fibonacci number defined by
$F_1=F_2=1$ and $F_n+F_{n+1}=F_{n+2}$ for any $n \geq 1$) instead of
$2^L$~\cite{pxp1,pxp2}. Thus, the Hilbert space dimension scales as
$\varphi^L$ for $L \gg 1$ where $\varphi=(1+\sqrt{5})/2$. For
completeness, we also write the Hamiltonian for the PXP model below:
\begin{eqnarray}
H_{\mathrm{PXP}} &=& \sum_{j=1}^L \tilde{\sigma}_j^x = \sum_{j=1}^L
P^{\downarrow}_{j-1} \left( |\uparrow_{j}\rangle \langle
\downarrow_{j} | +\mathrm{H.c.}\right) P^{\downarrow}_{j+1}.
\nonumber\\
  \label{Hpxpdef}
\end{eqnarray}
The $H_3$ interaction naturally arises when a Floquet version of the
PXP model is considered~\cite{dyn1,dyn5} with the time-dependent
Hamiltonian being
\begin{eqnarray}
H(t) &=&  w H_{\mathrm{PXP}}+h(t)\sum_i \sigma^z_j \label{fl1}
\end{eqnarray}
where $h(t)$ is a periodic function in time (with period $T$)
defined as $h(t)=-h_0$ for $0< t <T/2$ and $h(t)=+h_0$ for
$T/2<t<T$. For $|h_0| \gg |w|$, the Floquet Hamiltonian $H_F$
defined by $U(T)=\exp(-iH_FT/\hbar)$, where $U(T)$ denotes the time
evolution operator for one drive cycle, can be calculated using a
Floquet perturbation theory approach~\cite{fpt1,fpt2,fpt3} which
yields a power series of the form
\begin{eqnarray}
H_F &=& (w
f_1(h_0,T)+\cdots)H_{\mathrm{PXP}}+w^3g_3(h_0,T)H_3+\cdots
\nonumber\\ \label{fptfl}
\end{eqnarray}
with $H_3$ (Eq.~\ref{H3def}) being the leading non-PXP term and the
ellipsis indicating terms which either are higher order in $w/h_0$
or contain higher number of spins (for the exact form of
$f_{1}(\lambda,T)$ and $g_3(\lambda,T)$, see
Ref.~\onlinecite{dyn5}). Interestingly, all the terms
  with higher number of spins in Eq.~\ref{fptfl}
  like $H_5$, $H_7$, $\cdots$, $H_m$ (where $m$ is always
  odd)
  can be interpreted as Hamiltonians composed of purely off-diagonal
  terms connecting alternating spins on $m$ consecutive sites (e.g.,
  $\downarrow_j \uparrow \downarrow \uparrow \downarrow_{j+4}$ for $m=5$)
  with their
  flipped version
  ($\uparrow_j \downarrow \uparrow \downarrow \uparrow_{j+4}$ for $m=5$)
  with $P^\downarrow$ projectors at both the neighboring sites of the
  $m$-site units ($j-1,j+5$ for
  $m=5$) to ensure that
  the quantum dynamics respects the constrained nature of the Hilbert
  space. All such
  Hamiltonians, which can be viewed as generalized P${\mathbf{X}}_m$P
  models with an
  enlarged ${\mathbf{X}}_m$ that acts on $m \ge 3$ adjacent spins,
  lead to Hilbert
  space
  fragmentation on their own and we will focus on the most local one
  (in space) with
  $m=3$ in this work.

Both $H_3$ (Eq.~\ref{H3def}) and $H_{\mathrm{PXP}}$ (Eq.~\ref{Hpxpdef}) have
identical global symmetries on a finite lattice with periodic boundary
conditions. Both models have translational symmetry and a discrete spatial
inversion symmetry $I$ which maps a site $j$ to $L-j+1$. In addition, both the
Hamiltonians anticommute with the operator
\begin{eqnarray}
  Q=\prod_{j=1}^L \sigma_j^z
  \label{chiral}
  \end{eqnarray}
which leads to a particle-hole symmetry in the spectrum~\cite{pxp1,pxp2},
i.e., every eigenstate at energy $E$ (denoted by a state $|E\rangle$) has a
unique partner at energy $-E$ (the state $Q|E\rangle$). This
  particle-hole symmetry also forbids any $H_m$ with even $m$ in
  Eq.~\ref{fptfl}.

\begin{figure}[!htb]
  \includegraphics[width=0.96\hsize]{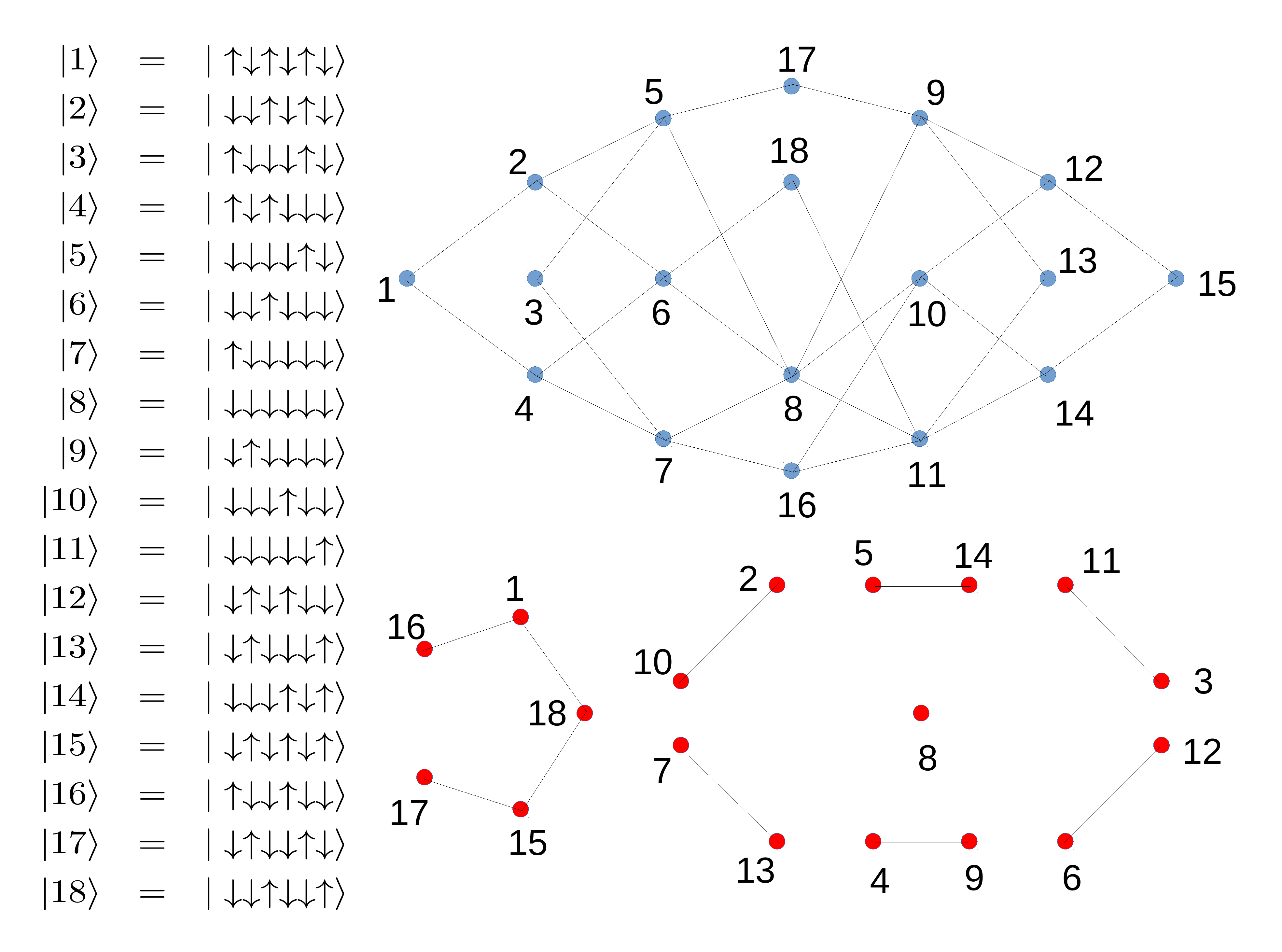}
  \caption{The $F_5+F_7=18$ Fock states for $L=6$ and their numbering shown in
    left panel. The right panel shows the connectivity graphs of
    $H_{\mathrm{PXP}}$ (in blue) and $H_3$ (in red) respectively.}
  \label{fig1}
\end{figure}

The key difference between the two models, $H_3$ and
  $H_{\mathrm{PXP}}$,
lies in the connectivity
graph induced by the interaction Hamiltonian in the computational
basis. Let us demonstrate this explicitly for a small system of
$L=6$ which has $F_5+F_7=18$ Fock states in the $S^z$ basis
(Fig.~\ref{fig1}). The connectivity graph then comprises of vertices
and links, where the vertices denote these Fock states and links are
placed between vertices that can be connected by a single action of
the Hamiltonian. While $H_{\mathrm{PXP}}$ produces an {\it ergodic}
graph where there exists at least one path between any two vertices,
the situation is markedly different for $H_3$ where the graph breaks
up into {\it dynamically disconnected} sectors (one lone vertex, six
pairs of vertices connected by a single link each and five vertices
connected by four links).

\begin{figure}[!htb]
  \includegraphics[width=0.96\hsize]{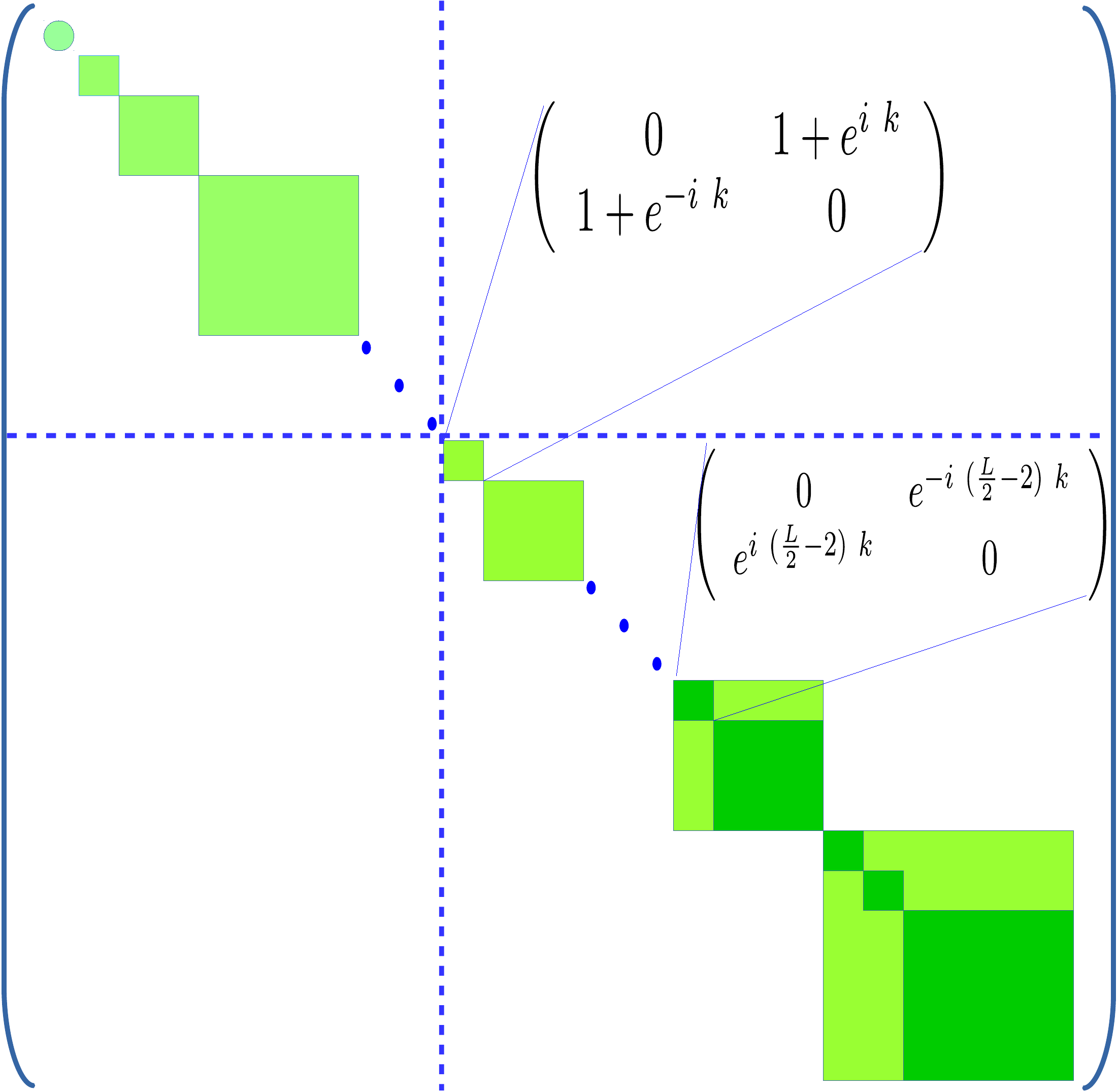}
  \caption{The Hamiltonian structure and the associated fragmentation is schematically shown here. The fragments in the top left quadrant (quadrants indicated
    by the blue dotted lines) are bubble type fragments whose sizes range from
    $1 \times 1$ to $2^{n_0} \times 2^{n_0}$ where $n_0 \sim L/5$. Each of
    these bubble fragments can be completely diagonalized due to the
    strictly localized nature of quasiparticles in such states.
    The non-bubble fragments
    in the bottom right quadrant are more appropriately expressed in the momentum
    basis. A $2 \times 2$ non-bubble fragment (with the
    explicit expression for $H_3$ shown) is indicated. Some
    bigger non-bubble fragments (indicated in light green) split up into
    smaller secondary fragments (indicated in dark green) after a further
    basis transformation involving a fixed number of basis states. Some of
    these secondary fragments, therefore, have a fixed dimension with increasing
    system size. One such
    $2 \times 2$ secondary fragment and the corresponding expression of $H_3$
    is indicated above.}
  \label{fig2}
\end{figure}

These dynamically disconnected sectors lead to a rich fragmented
structure of the Hilbert space for $H_3$ as we will describe in the
rest of the paper. The fragmentation of the Hamiltonian is
schematically shown in Fig.~\ref{fig2} where the fragments in the
top left quadrant (quadrants marked by blue dotted lines) have
dimension $2^{n} \times 2^{n}$ with $n$ being an integer and ranging
from $1$ to $n_0 \sim L/5$. These fragments represent bubble states
with strictly localized quasiparticles (hence, a real space
description being appropriate). Due to the strict localization
property, all eigenstates of any bubble type fragment of size $2^n
\times 2^n$ can be computed. The bottom right quadrant shows some of
the analytically tractable states with mobile quasiparticles that
are more appropriately written in the momentum basis. In the bottom
right quadrant (Fig.~\ref{fig2}), we have schematically shown an
example of primary fragmentation in momentum space which leads to a
$2 \times 2$ matrix (indicated in light green with the corresponding
matrix for $H_3$ written out).

We have also schematically shown examples of a secondary
fragmentation mechanism due to specific linear combinations of the
basis states whereby a primary fragment (indicated by light green)
splits up into two or more secondary fragments (indicated by dark
green). Some of these secondary fragments have a
fixed dimension (e.g., a $2 \times 2$ example shown in
Fig.~\ref{fig2} with the corresponding matrix for $H_3$ explicitly
indicated) that does not scale with increasing system size. This
feature makes them analytically tractable even for large $L$. The
non-bubble fragments whose dimensions scale with the system size do
not seem to have any simple representation of the corresponding
high-energy eigenstates in terms of quasiparticles; it is likely
that such states satisfy ETH when $L \gg 1$.

\begin{widetext}
\begin{figure*}[!tbp]
  \includegraphics[width=0.49\hsize]{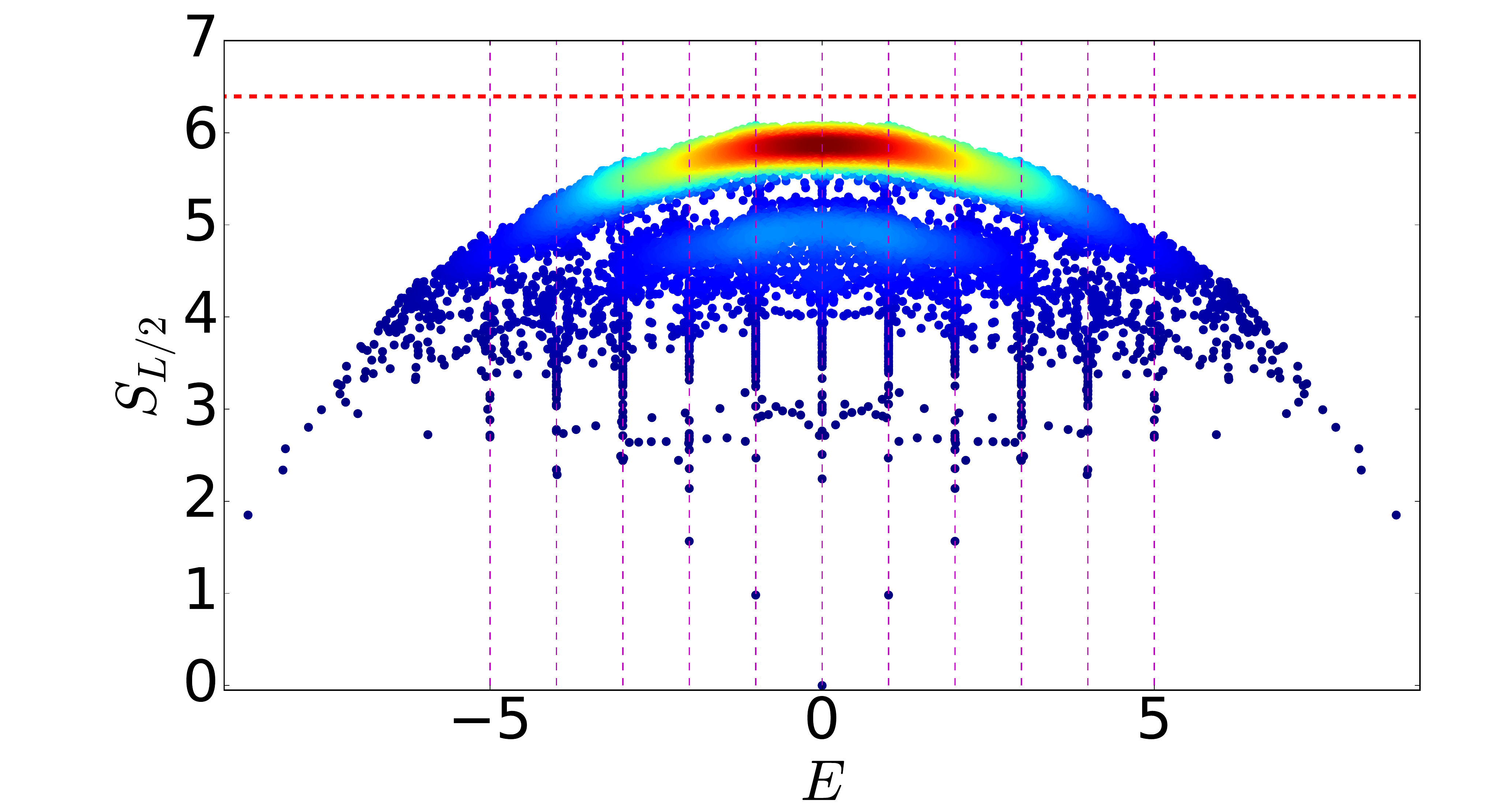}%
  \includegraphics[width=0.49\hsize]{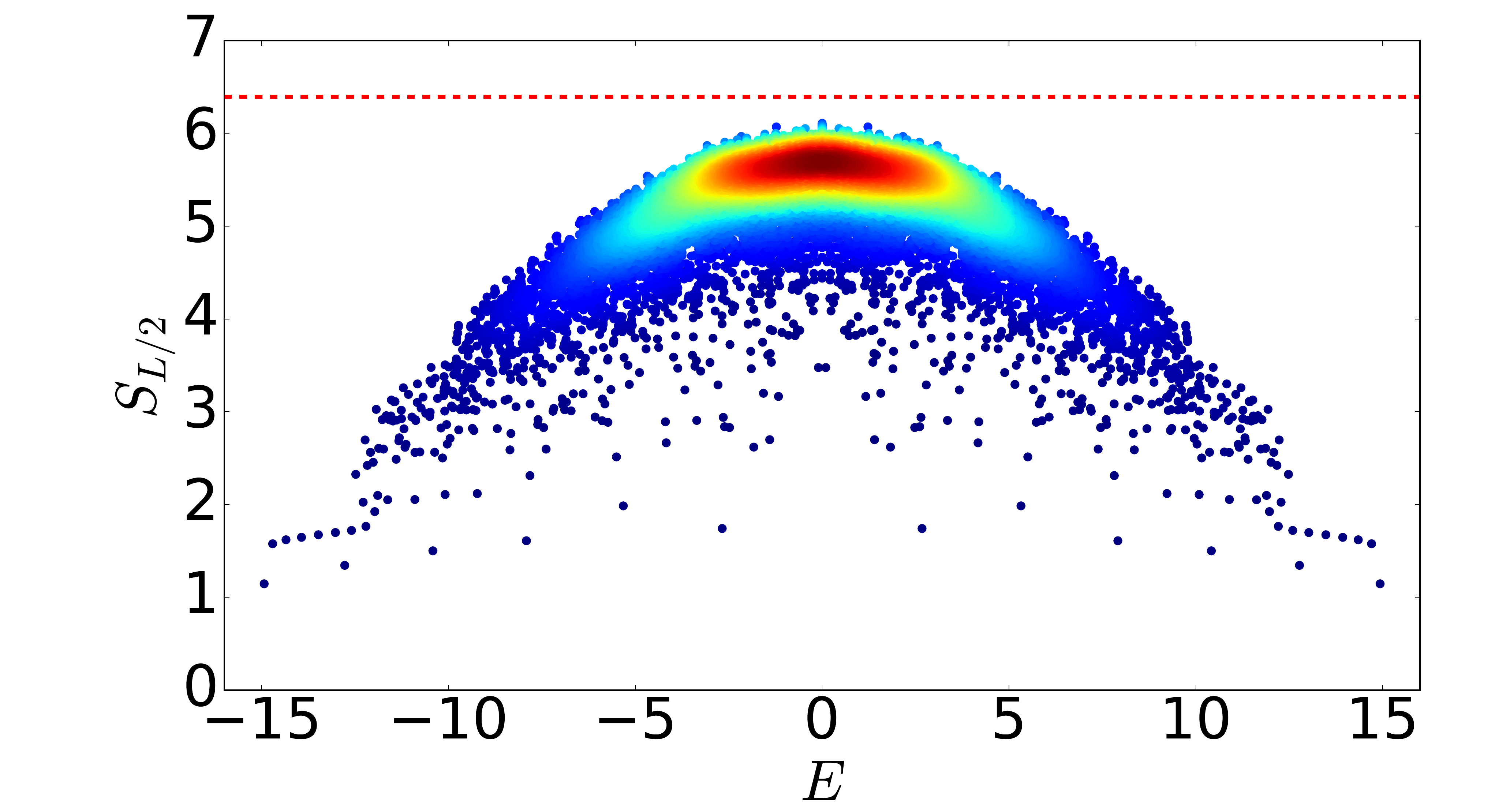}\\
  \includegraphics[width=0.49\hsize]{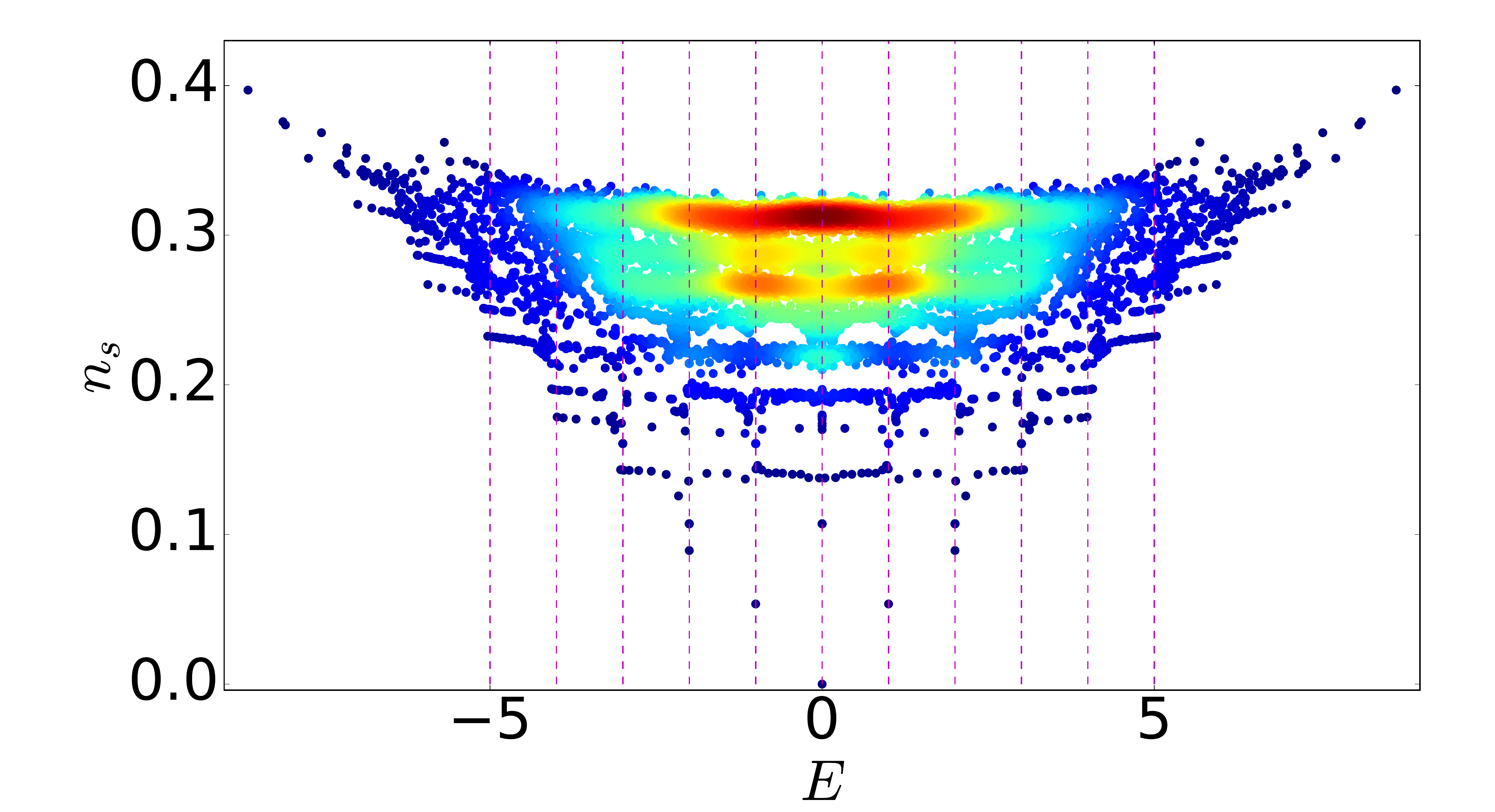}%
   \includegraphics[width=0.49\hsize]{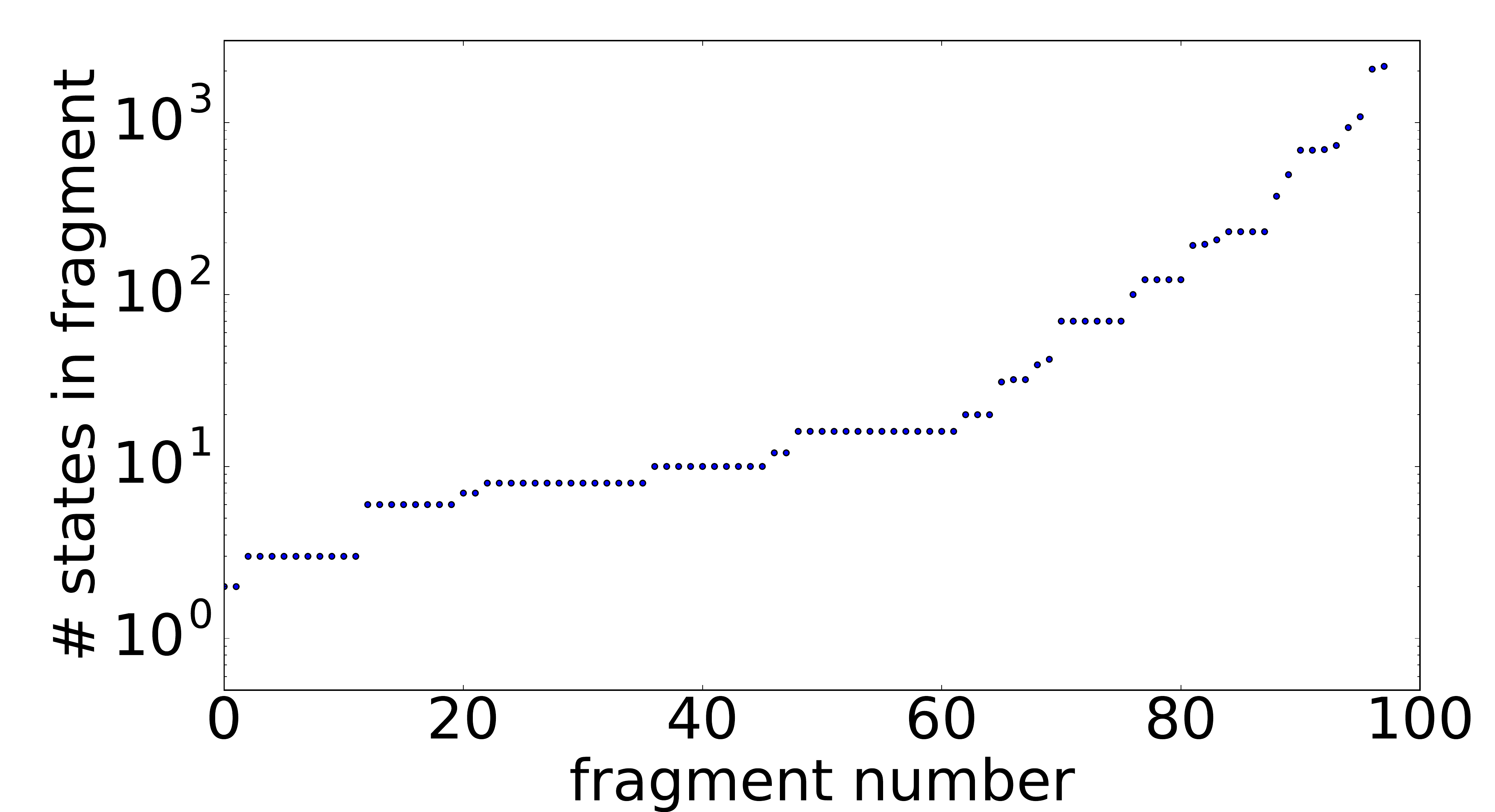}
  \caption{The bipartite entanglement entropy with equal partitions, $S_{L/2}$,
shown for all energy eigenstates in the sector with zero momentum
and spatial inversion symmetry for $L=28$ in the top panels.
The top left panel shows data for $H_3$ while the top right
panel correspond to $H_{\mathrm{PXP}}$.
  The horizontal dotted line indicates the average entanglement entropy,
  $S_{\mathrm{Page}}$, of random pure states in both the top panels.
The bottom left panel shows
the expectation value of the operator $n_s=\sum_{j=1}^L
(\sigma_j^z+1)/(2L)$ for the eigenstates of $H_3$. The density of
states is indicated by the same color map in all three panels
where warmer color corresponds to higher density of states.
Additionally, dotted vertical lines indicate integer
values of $E$ for the top left and bottom left panels. The bottom
right panel shows the number of primary fragments induced by $H_3$
and their dimensions for the zero momentum and spatial inversion
symmetry for $L=28$. }
\label{fig3}
\end{figure*}
\end{widetext}

To further contrast the difference between the spectrum of $H_3$
(Eq.~\ref{H3def}) and $H_{\mathrm{PXP}}$ (Eq.~\ref{Hpxpdef}), we
calculate the bipartite entanglement entropy $S_{L/2}$  for $L=28$
in the sector with zero momentum ($k=0$) and spatial inversion
symmetry ($I=+1$) for both the models. The bipartite
entanglement entropy is given by
\begin{eqnarray}
  S_{L/2}=-\mathrm{Tr}[\rho_A \ln \rho_A]
  \label{reduced}
\end{eqnarray}
for each eigenstate $|\Psi\rangle$ where
$\rho_A=\mathrm{Tr}_B|\Psi\rangle \langle \Psi|$ represents the
reduced density matrix. Here $\rho_A$ is computed by partitioning
the one-dimensional chain into two equal halves $A$ and $B$. While
typical high-energy eigenstates are expected to show a volume law
scaling of $S_{L/2} \sim L$, anomalous high-energy eigenstates
usually have much smaller values of $S_{L/2}$. While the bipartite
entanglement entropy for $H_{\mathrm{PXP}}$ (Fig.~\ref{fig3} (top
right panel)) shows the characteristic scar states of the PXP model
as outliers of $S_{L/2}$ that are almost equally spaced in energy,
its behavior for $H_3$ (Fig.~\ref{fig3} (top left panel)) is
markedly different. The latter shows the presence of many more
anomalous eigenstates at both integer as well as non-integer
eigenvalues. In fact, an entanglement gap seems to emerge in
Fig.~\ref{fig3}(top left panel) which separates the low and high
entanglement entropy eigenstates and this gap should become more
pronounced with increasing $L$.
It is also clear by comparing both
the top panels in Fig.~\ref{fig3} that just like the PXP model, the
bulk of the states have large volume-law entanglement in the
spectrum of $H_3$ (the density of states is shown using the same
color map in both panels where warmer color indicates higher
density of states) indicating the non-integrable nature of the
model. In fact, $S_{L/2}$ approaches a value close to
  the average entanglement entropy,
  $S_{\mathrm{Page}}=\ln (\mathcal{D}_{L/2}) -\frac{1}{2}$ of
  pure random states~\cite{Page}
  where $\mathcal{D}_{L/2}$ is the Hilbert space dimension of A (B), for the
mid-spectrum eigenstates in both the top panels of Fig.~\ref{fig3}.

The bottom left panel of Fig.~\ref{fig3} shows the expectation value of the
operator $n_s=\sum_{j=1}^L (\sigma^z_j+1)/(2L)$ for all the eigenstates of
$H_3$ at
$L=28$ with $k=0$ and $I=+1$ which again highlights that several
of the high-energy eigenstates do not seem to satisfy ETH. The bottom
right panel of Fig.~\ref{fig3} shows the variations in the sizes of the
different (primary) fragments that arise for this symmetry sector with the
smallest fragment being $1 \times 1$ and the largest being $2125\times 2125$.
Lastly, analyzing the data for both the models at $L=28$
  with $k=0, I=+1$ clearly shows that while $H_{\mathrm{PXP}}$ does not show
  any residual degeneracy in the spectrum apart from the zero modes, the
  spectrum of $H_3$ shows exact degeneracies even at non-zero integer values
  (indicated by dotted vertical lines in Fig.~\ref{fig3} (top left and bottom
  left
  panels)).

\section{Construction of bubble eigenstates}
\label{sec:bubblestates} There is a simple class of exact
eigenstates that we can write explicitly for $H_3$ for any $L$,
where these have integer energies and also satisfy strict area law
scaling of entanglement entropy in the thermodynamic limit. Let us
start with the Fock state with all $S^z=\downarrow$. This state is
annihilated by $H_3$ and is, therefore, an ``inert'' state which
forms a $1\times 1$ fragment in the Hilbert space. Interestingly,
while earlier models of fragmentation contained an exponentially
large number of inert states for a given system size, this
one-dimensional model only contains one such state.

Taking the inert state as a reference state, let us now choose a
arbitrary site $i$ and change the corresponding spin to
$S_i^z=\uparrow$. Importantly, this particular Fock state is only
connected to one other Fock state in the Hilbert space under the
action of $H_3$ leading to a $2 \times 2$ fragment. Looking at both
the Fock states, only the $(i-1,i,i+1)$ spins can differ from the
reference state and we denote these two types of building blocks of
three adjacent spins as
\begin{eqnarray}
\boxed{X_{1,i}} &=& \boxed{\uparrow_{i-1} \downarrow_i
\uparrow_{i+1}} \nonumber\\
\boxed{X_{2,i}} &=& \boxed{\downarrow_{i-1} \uparrow_i
  \downarrow_{i+1}}
\label{bubstates}.
\end{eqnarray}
These blocks are immersed in the background of the rest of the
spins being all $\downarrow$ as provided by the reference inert
state. Two such states connected under the action of $H_3$ are shown
below:
\begin{eqnarray}
  \cdots \downarrow \downarrow \downarrow \boxed{X_{1,i}}\downarrow \downarrow \downarrow \downarrow \downarrow \downarrow \cdots  \nonumber \\
  \cdots \downarrow \downarrow \downarrow \boxed{X_{2,i}}\downarrow \downarrow \downarrow \downarrow \downarrow \downarrow \cdots
\label{onebubbleFock}
\end{eqnarray}

Placing $n$ such $X_{1,2}$ units, where each of the units is
``sufficiently distant'' from the others, in the background of the
inert state then shows that this Fock state is connected to $2^n-1$
other Fock states under $H_3$ where any of the $X_{1}$ unit can be
changed to a $X_{2}$ unit and vice-versa. The question is how close
can these units be for this strict local picture of $X_1
\leftrightarrow X_2$ conversions under $H_3$ to hold. For this, it
is enough to consider the case where there are two such units
inserted in the reference state. Such a consideration
shows that if a minimum of two adjacent $S^z=\downarrow$ separate
the two $X_{1,2}$ units, then the Hilbert space fragment continues
to be $4 \times 4$ with the four Fock states simply being all
combinations of $(X_{1(2),i},X_{1(2),j})$ where
$i,j$ denote the central sites of the two units respectively, as
shown below:
\begin{eqnarray}
  \cdots \downarrow \downarrow \downarrow \boxed{X_{1,i}}\downarrow \downarrow \boxed{X_{1,j}} \downarrow \downarrow \cdots  \nonumber \\
  \cdots \downarrow \downarrow \downarrow \boxed{X_{1,i}}\downarrow \downarrow \boxed{X_{2,j}} \downarrow \downarrow \cdots \nonumber \\
  \cdots \downarrow \downarrow \downarrow \boxed{X_{2,i}}\downarrow \downarrow \boxed{X_{1,j}} \downarrow \downarrow \cdots  \nonumber \\
  \cdots \downarrow \downarrow \downarrow \boxed{X_{2,i}}\downarrow \downarrow \boxed{X_{2,j}} \downarrow \downarrow \cdots
  \label{twobubbleFock}
  \end{eqnarray}
Thus, bubble Fock states are constructed of various $X_{1(2)}$
immersed in a background of the rest of the spins being
$S^z=\downarrow$ with the additional constraint that no two
$X_{1(2)}$ units have less than two adjacent $\downarrow \downarrow$
separating them. Due to the hard core constraint of at least two
consecutive $\downarrow$ spins between any two $X_{1,2}$ units, the
maximum value of $n$ (total number of $X_{1,2}$ units in a bubble
Fock state) equals $n_0=\lfloor L/5 \rfloor$ where $\lfloor \mbox{~}
\rfloor$ denotes the floor function. Thus, for $L \gg 1$, the
largest bubble-type Hilbert space fragment equals $2^{L/5} \times
2^{L/5}$.

Given that $H_3$ simply converts a $X_{1,j}$ to a $X_{2,j}$ and
vice-versa for bubble Fock states, it is straightforward to
diagonalize each of these fragments with dimension $2^n \times 2^n$
where $n=1,2,\cdots,n_0$. Denoting $X_{1,j}$ and $X_{2,j}$ as
pseudospin-$1/2$ variables
  with
  $\tau_j^z=+1$ and $\tau_j^z=-1$, the form of $H_3$ projected to
  any $n \neq 0$ bubble type fragment simply equals
\begin{eqnarray}
  H_{3,\mathrm{bubble}} = \sum_{j_b=1}^{n} \tau_{j_b}^x
  \label{bubbleHfree}
\end{eqnarray}
where $j_b$ denotes the central sites of the bubbles. This
``non-interacting'' effective Hamiltonian can be diagonalized for
any $n$ to give that a quantum state build out of $n_1$
$\boxed{\downarrow_{j-2}X_j \downarrow_{j+2}}$ units, $n_2$
$\boxed{\downarrow_{j-2}Y_j \downarrow_{j+2}}$ units (where
$n_1+n_2=n$) and the rest being $S^z=\downarrow$, where
\begin{eqnarray}
X_j &=& \frac{1}{\sqrt{2}} (X_{1,j}+X_{2,j}), \quad {\rm and} \quad
Y_j = \frac{1}{\sqrt{2}} (X_{1,j}-X_{2,j}) \nonumber\\ \label{xydef}
\end{eqnarray}
are exact eigenstates of $H_3$ with eigenvalues $E=n_1-n_2$. Note
that fixing two $\downarrow$ at both sides of $X_j$ and $Y_j$
automatically ensures that two such units are separated by at least
two consecutive $\downarrow$ spins, as required for bubble Fock
states. E.g., the four eigenstates obtained by diagonalizing the $4
\times 4$ matrix obtained for $n=2$ are as follows:
\begin{eqnarray}
  &&\cdots \downarrow \downarrow \otimes \boxed{\downarrow_{j-2}X_j \downarrow_{j+2}} \otimes \boxed{\downarrow_{j+3}X_{j+5} \downarrow_{j+7}}\otimes \downarrow \cdots \nonumber \\
  &&\cdots \downarrow \downarrow \otimes \boxed{\downarrow_{j-2}X_j \downarrow_{j+2}} \otimes \boxed{\downarrow_{j+3}Y_{j+5} \downarrow_{j+7}}\otimes \downarrow \cdots  \nonumber \\
  &&\cdots \downarrow \downarrow \otimes \boxed{\downarrow_{j-2}Y_j \downarrow_{j+2}} \otimes \boxed{\downarrow_{j+3}X_{j+5} \downarrow_{j+7}}\otimes \downarrow \cdots \nonumber \\
  &&\cdots \downarrow \downarrow \otimes \boxed{\downarrow_{j-2}Y_j \downarrow_{j+2}} \otimes \boxed{\downarrow_{j+3}Y_{j+5} \downarrow_{j+7}}\otimes \downarrow \cdots \nonumber \\
  \end{eqnarray}

Thinking of $X_j, Y_j$ as local quasiparticles of $H_3$, the bubble
eigenstates have the property that they are composed of strictly
localized or immobile quasiparticles and have integer energies. From
the construction of the eigenstates, it is evident that these
satisfy an area law for entanglement entropy where the bipartite
entanglement entropy can only be $S_{L/2}=0$, $\ln 2$ or $2 \ln 2$ depending
on whether the bipartition intersects zero, one or two $X/Y$ units
in spite of being high-energy eigenstates. For example, the bubble
eigenstates at $E=0$ are mid-spectrum states owing to the $E$ to
$-E$ symmetry of the spectrum of $H_3$ and, therefore, violate ETH
since such states are expected to have volume law scaling of
entanglement entropy.

The number of bubble eigenstates at an integer energy $E=n_1-n_2$
can be calculated for a given system size
from the number of configurations that satisfy the hard core constraint of
at least two consecutive $\downarrow$ spins separating any of the $n_1$ $X$
and $n_2$ $Y$ units on the periodic one-dimensional lattice. In
Table.~\ref{table1}, we compare the number of bubble eigenstates at each
$E=0, \pm 1, \pm 2, \cdots$ (the number of bubble eigenstates for $E=n$
equals that for $E=-n$ because of the $E$ to $-E$ symmetry of the spectrum)
for chains with $L=16,18,20$ with the degeneracy of integer eigenvalues
obtained from the exact diagonalization data (ED) (without restricting to any
symmetry sector). Comparing the degeneracy of the bubble eigenstates with the
integer eigenstates, we see that for $L=16,18,20$, $E=\pm n_0$ eigenstates
are solely the bubble eigenstates. We conjecture that this property
is true for any $L$. In fact, for $L=20$, even the $E=\pm (n_0-1)$ eigenstates
are all bubble eigenstates. However, the situation is quite different as we
move closer to $E=0$ where it is clear that there are a large number of
non-bubble eigenstates that have integer eigenvalues as well.
\begin{table}
  \caption{The number of bubble eigenstates at integer energies and the number of integer eigenstates obtained from exact diagonalization for chain lengths $L=16,18,20$.}
  \label{table1}
\begin{center}
  \begin{tabular}{|c|c|c|}
    \hline \hline
    $L=16$ & Bubble eigenstates & Exact Diagonalization \\
    \hline \hline
    $E=+3$ (or $-3$) & $16$ & $16$ \\
    \hline
    $E=+2$ (or $-2$) & $56$ & $57$ \\
    \hline
    $E=+1$ (or $-1$) & $64$ & $132$ \\
    \hline
    $E=0$ & $113$ & $193$ \\
    \hline \hline
    $L=18$ & Bubble eigenstates & Exact Diagonalization \\
    \hline \hline
    $E=+3$ (or $-3$) & $60$ & $60$ \\
    \hline
    $E=+2$ (or $-2$) & $81$ & $100$ \\
    \hline
    $E=+1$ (or $-1$) & $198$ & $308$ \\
    \hline
    $E=0$ & $163$ &  $348$ \\
    \hline \hline
    $L=20$ & Bubble eigenstates & Exact Diagonalization \\
    \hline \hline
     $E=+4$ (or $-4$) & $5$ & $5$ \\
    \hline
    $E=+3$ (or $-3$) & $140$ & $140$ \\
    \hline
    $E=+2$ (or $-2$) & $130$ & $255$ \\
    \hline
     $E=+1$ (or $-1$) & $440$ & $600$ \\
    \hline
    $E=0$  &$251$ & $709$ \\
    \hline
    \end{tabular}
  \end{center}
\end{table}

While we have not been able to find a closed form expression for the
degeneracy of the different bubble eigenstates for an arbitrary $L$, we can
still place a lower bound on the degeneracy of such states when $L \gg 1$.
Consider a chain of length $L=10m$ where $m$ is an integer. Then, the maximum
number of $X/Y$ units that can be accommodated due to the hard core
constraints equals $n_0=2m$. Ignoring whether a bubble is $X/Y$ type, there
are only $5$ distinct ways of placing all the $n_0$ bubbles in the chain
since each bubble is separated from the neighboring ones by precisely
two lattice sites when $L=10m$. To construct a bubble eigenstate with energy
$E=2n-n_0$, $n$ bubbles need to be made $X$ type and the rest ($n_0-n$) $Y$
type which introduces an additional degeneracy of ${n_0 \choose n}$. Assuming
both $n_0,n \gg 1$, we then get that the degeneracy of bubble eigenstates with
energy $E=2n-n_0$ (denoted by $\Omega_B(n)$ below) is bounded below by
\begin{eqnarray}
  \Omega_B(n) > 5 \times 2^{n_0}\sqrt{\frac{2}{\pi n_0}} \exp \left( 2n_0 \left(x-\frac{1}{2}\right)^2\right)
\label{bubblebound}
\end{eqnarray}
where $n_0=L/5$ and $x=n/n_0$ for $L \gg 1$. This bound immediately
shows that the
number of bubble eigenstates is exponentially large in the system size
for integer energies that range from $E=0$ to $|E| \sim \mathrm{O}(\sqrt{L})$
while the
maximum value of the integer energy $|E| \sim L/5$ when $L \gg 1$. Since the
bubble eigenstates satisfy area law scaling of entanglement entropy (with a
maximum $S_{L/2}=2\ln(2)$), it is an interesting open question to find
how the number of
non-bubble integer eigenstates scales with system size and how many of such
eigenstates satisfy volume law scaling of entanglement entropy, especially
in the neighborhood of $E=0$, given the non-integrable nature of $H_3$ and
the numerical data presented in Fig.~\ref{fig3} (top left panel).

While the bubble eigenstates have strictly localized quasiparticles, there are
other anomalous high-energy excitations for this model which we collectively
dub as ``non-bubble'' eigenstates. These are necessarily composed of mobile
excitations and seem to have a more natural description in momentum space.
Unlike the bubble eigenstates, we do not claim to have a complete
understanding of these special eigenstates.
In the next two sections, we describe a
class of such non-bubble eigenstates of $H_3$.

\section{Non-bubble eigenstates with dispersing quasiparticles}
\label{sec:nonbubblestates}

\begin{figure}
 \includegraphics[width=0.96\hsize]{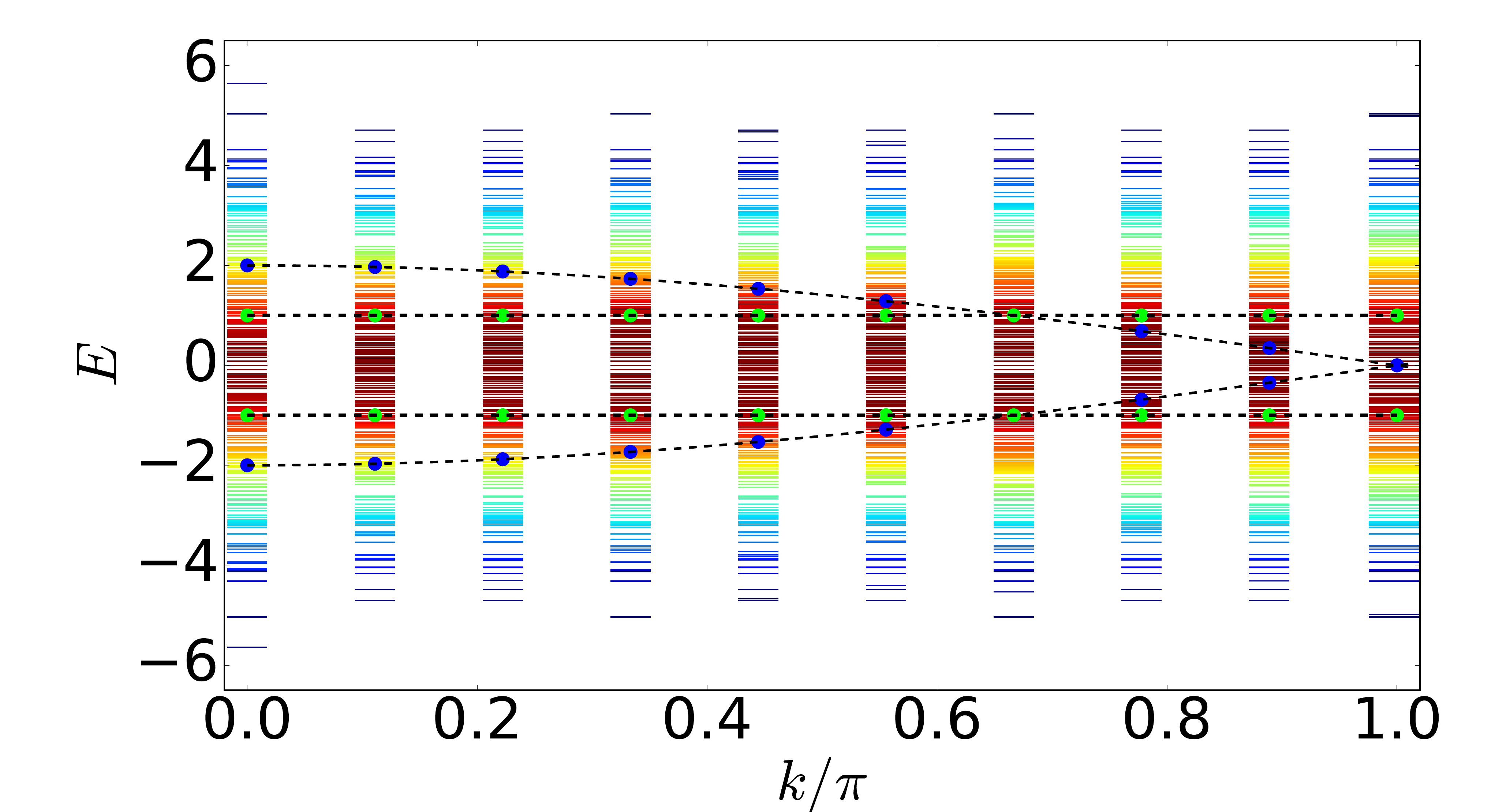}
 \caption{The momentum resolved energy spectrum
 for $L=18$ where the color map indicates the density
 of states. The non-bubble eigenstates that lead to
 two dispersive bands and two flat bands in momentum
 space respectively are highlighted as blue and green
 dots respectively. The dashed lines are a guide to the eye.
 The color map indicates the momentum resolved density of
 states. Warmer color corresponds to higher density of states.}
\label{fig4}
\end{figure}

The simplest non-bubble eigenstates are composed starting from Fock
states with two $X_{2}$ units next to each other with the rest of
the spins being $S^z=\downarrow$ provided by the (inert) reference
state. The action of $H_3$ then leads to an effective hopping of
these two $X_{2}$ units together either in the forward or backward
direction through intermediate Fock states with two consecutive
$X_1,X_2$ units that also move forward or backward in the same
manner.
\begin{widetext}
\begin{eqnarray}
  \cdot \cdot \downarrow \downarrow \boxed{X_{2,j}}\boxed{X_{2,j+3}} \downarrow \downarrow \cdot \cdot \leftrightarrows \cdot \cdot \downarrow \downarrow \boxed{X_{1,j}}\boxed{X_{2,j+3}}  \downarrow \downarrow \cdot \cdot \leftrightarrows \cdot \cdot  \downarrow \boxed{X_{2,j-1}}\boxed{X_{2,j+2}} \downarrow \downarrow \downarrow \cdot \cdot \leftrightarrows \cdot \cdot  \downarrow \boxed{X_{1,j-1}}\boxed{X_{2,j+2}}  \downarrow \downarrow \downarrow \cdot \cdot \nonumber \\
  \cdot \cdot \downarrow \downarrow \boxed{X_{2,j}}\boxed{X_{2,j+3}} \downarrow \downarrow \cdot \cdot \leftrightarrows \cdot \cdot  \downarrow \downarrow \boxed{X_{2,j}}\boxed{X_{1,j+3}}  \downarrow \downarrow \cdot \cdot \leftrightarrows \cdot \cdot \downarrow \downarrow \downarrow \boxed{X_{2,j+1}}\boxed{X_{2,j+4}}  \downarrow \cdot \cdot \leftrightarrows \cdot \cdot  \downarrow \downarrow \downarrow \boxed{X_{2,j+1}}\boxed{X_{1,j+4}}  \downarrow  \cdot \cdot \nonumber \\
  \label{mobileFock}
\end{eqnarray}
\end{widetext}
Using this observation, we find it convenient to go to momentum space
and define the following two basis states:
\begin{eqnarray}
  |\Phi_{1,k}\rangle &=& \frac{1}{\sqrt{L}}\sum_{j=1}^L \exp(ikj)T^j |\uparrow \downarrow \downarrow \uparrow \underbrace{\downarrow \downarrow \cdots \downarrow \downarrow}_{L-4} \rangle \nonumber \\
  |\Phi_{2,k}\rangle &=& \frac{1}{\sqrt{L}}\sum_{j=1}^L \exp(ikj)T^j |\uparrow \downarrow \uparrow \downarrow \uparrow \underbrace{\downarrow \cdots \downarrow \downarrow}_{L-5} \rangle
  \label{fragmentink}
  \end{eqnarray}
where $T$ represents a global translation by one lattice unit and the momentum
$k=2\pi n/L$ where $n=0,1,2,\cdots,L-1$. The spins in $\underbrace{\downarrow
  \cdots \downarrow}$ are all assumed to be $\downarrow$. These basis states
then define a $2 \times 2$ fragment in momentum space at
each $k$ because
\begin{eqnarray}
  H_3|\Phi_{1,k}\rangle=(1+\exp(ik))|\Phi_{2,k}\rangle.
  \label{2times2k}
  \end{eqnarray}
Diagonalizing the $2\times 2$ matrix leads to the following
exact eigenenergies and eigenvectors:
\begin{eqnarray}
E_{\pm} (k) &=& \pm 2 \cos k/2 \nonumber\\
|\Psi_{\pm}(k) \rangle &=& \frac{\exp(-ik/2)|\Phi_{1,k}\rangle \pm
|\Phi_{2,k}\rangle}{\sqrt{2}}
  \label{dispersingnonbubbles}
  \end{eqnarray}
This leads to two dispersive bands composed of a total of $2L$
anomalous eigenvectors at high energies. Thus, we have obtained a
very simple class of exact eigenstates with irrational energies in
general for any $L \gg 1$. E.g., for any $L \geq 8$ which is a
multiple of four, such that $k=\pi/2$ is allowed, these eigenstates
have energies $E=\pm \sqrt{2}$ at that particular momentum. In fact,
for $L=16$, such a non-bubble eigenstate with $E=2(-2)$ state at
$k=0$ provides the missing eigenstate apart from the $56$ bubble
eigenstates when we compare with the ED results
(Table.~\ref{table1}). Moreover, these states have
the symmetry $E_{\pm}(2\pi-k)= E_{\mp}(k)$. Thus they intersect at
$k= \pi$ leading to two zero energy non-bubble eigenstates. In
Fig.~\ref{fig4}, we show the presence of these two dispersive bands
of eigenstates in momentum space for a chain of $L=18$ after
resolving the ED data in momentum $k$.

\section{Secondary fragmentation of the Hilbert space}
\label{sec:secondaryfrag} We now point out a novel secondary
fragmentation mechanism, which did not appear in previous models to
the best of our knowledge, whereby certain specific linear
combinations of a fixed number of basis states contained in a large
primary fragment (whose dimension keeps increasing with system size)
turns out to be orthogonal to all other vectors inside the fragment.
In addition, they also form a closed subspace under the action of
$H_3$ leading to a smaller emergent fragment of a fixed dimension in
the Hilbert space. We note that though most of the eigenvalues
inside the larger primary fragments are irrational in nature, as
expected for a non-integrable model, diagonalization of these
smaller secondary fragments gives integer eigenvalues. These
non-bubble states with integer eigenvalues exist inside the larger
primary fragments of all momentum sectors and their number increases
with increasing system size. We discuss two such examples of exact
non-bubble eigenstates that emerge from this mechanism below.

\subsection{Secondary fragmentation in $k=0, I=+1$}
\label{subsec:k0I1}
We first discuss the formation of an exact state inside the largest
primary fragment of $k=0, I=+1$. We note here that the state
$(|{\mathbb Z}_2\rangle + |{\bar{\mathbb Z}}_2\rangle)/\sqrt{2}$
belongs to this fragment for $L=10,16,22,28\cdots$; however for
other $L$, this is not the case. To this end we consider the
following four basis states with $k=0$, $I=+1$ defined as
\begin{eqnarray}
  |v_1\rangle &=& \frac{1}{\sqrt{2L}} \sum_{j=1}^L T^j (1+I) |\uparrow \downarrow \downarrow \uparrow \downarrow \uparrow \downarrow \downarrow \downarrow \uparrow
  \underbrace{\downarrow \cdots \downarrow}_{L-10}\rangle \nonumber \\
  |v_2\rangle &=& \frac{1}{\sqrt{2L}} \sum_{j=1}^L T^j (1+I) |\uparrow \downarrow \uparrow \downarrow \downarrow \uparrow \downarrow \downarrow \downarrow \downarrow \uparrow \underbrace{\downarrow \cdots \downarrow}_{L-11}\rangle \nonumber \\
  |v_3\rangle &=& \frac{1}{\sqrt{2L}} \sum_{j=1}^L T^j (1+I) |\uparrow \downarrow \uparrow \downarrow \downarrow \uparrow \downarrow \uparrow \downarrow \downarrow \uparrow \underbrace{\downarrow \cdots \downarrow}_{L-11}\rangle \nonumber \\
  |v_4\rangle &=& \frac{1}{\sqrt{2L}} \sum_{j=1}^L T^j (1+I)|\uparrow \downarrow \uparrow \downarrow \downarrow \uparrow \downarrow \downarrow \downarrow \uparrow \downarrow \uparrow \underbrace{\downarrow \cdots \downarrow}_{L-12}\rangle \nonumber \\
  \label{secfrag1}
  \end{eqnarray}
Remarkably, the following linear combinations
\begin{eqnarray}
  |\psi_a\rangle &=& \frac{1}{\sqrt{2}} (|v_1\rangle -|v_2\rangle) \nonumber \\
  |\psi_b\rangle &=& \frac{1}{\sqrt{2}} (|v_3\rangle -|v_4\rangle)
  \label{secfrag2}
\end{eqnarray}
are orthogonal to the rest of the states in the fragment and are
closed under the action of $H_3$ since $H_3|\psi_{a,b}\rangle
=-|\psi_{b,a} \rangle$. To see why this is the case, we first note
that the action of $H_3$ on $|v_1\rangle$ and $|v_2\rangle$ leads to
states with three up-spins that are clearly outside the subspace
described above.  Similarly $H_3$ acting on $|v_3\rangle$ and
$|v_4\rangle$ leads to states with six up-spins. The particular
linear combinations chosen in Eq.\ \ref{secfrag2} cancel the
amplitude of these states. Moreover $H_3$ acting on $|v_1\rangle$
and $|v_2\rangle$ also generate five up-spin states other than
$|v_3\rangle$ and $|v_4\rangle$; the amplitude of these states also
cancel with the linear combination mentioned above. This leads to
this secondary fragmentation for any $L\geq 16$. Diagonalizing this
$2 \times 2$ fragment then gives the following eigenstates
\begin{eqnarray}
  \frac{1}{\sqrt{2}} (|\psi_a\rangle \pm |\psi_b\rangle )
  \label{secfrag3}
  \end{eqnarray}
with $E=\mp 1$.

\subsection{Non-bubble eigenstates with flat bands}
\label{subsec:flatbands}

We also find a class of high-energy eigenstates in this model
that form flat bands in momentum space. The relevant basis states in
momentum space are as follows
\begin{eqnarray}
  |w_1 \rangle &=& \frac{1}{\sqrt{L}}\sum_{j=1}^L \exp(ikj)T^j |\uparrow \underbrace{\downarrow \cdots \downarrow}_{\frac{L}{2}-4} \uparrow \downarrow \uparrow \downarrow \downarrow \uparrow  \underbrace{\downarrow \cdots \downarrow}_{\frac{L}{2}-3}\rangle \nonumber \\
  |w_2 \rangle &=& \frac{1}{\sqrt{L}}\sum_{j=1}^L \exp(ikj)T^j |\uparrow \downarrow \downarrow \uparrow \downarrow \uparrow  \underbrace{\downarrow \cdots \downarrow}_{\frac{L}{2}-4} \uparrow  \underbrace{\downarrow \cdots \downarrow}_{\frac{L}{2}-3}\rangle \nonumber \\
  |w_3 \rangle &=& \frac{1}{\sqrt{L}}\sum_{j=1}^L \exp(ikj)T^j |\uparrow \downarrow \uparrow \underbrace{\downarrow \cdots \downarrow}_{\frac{L}{2}-5} \uparrow \downarrow \uparrow \downarrow \downarrow \uparrow \underbrace{\downarrow \cdots \downarrow}_{\frac{L}{2}-4}\rangle \nonumber \\
  |w_4 \rangle &=& \frac{1}{\sqrt{L}}\sum_{j=1}^L \exp(ikj)T^j |\uparrow \downarrow \downarrow \uparrow \downarrow \uparrow \underbrace{\downarrow \cdots \downarrow}_{\frac{L}{2}-5} \uparrow \downarrow \uparrow \underbrace{\downarrow \cdots \downarrow}_{\frac{L}{2}-4} \rangle \nonumber \\
  \label{flatband1}
\end{eqnarray}
Note that $\langle w_1|H_3|w_3 \rangle =\exp(-ik)$ and $\langle
w_2|H_3|w_4 \rangle=1$. An appropriate linear combination of
$|w_i\rangle$ ($i=1,2$ and $i=3,4$) can be found following the same
principle described in the previous subsection; these linear
combinations are
\begin{eqnarray}
  |\chi_a\rangle &=& \frac{1}{\sqrt{2}} \left(|w_1\rangle -e^{-i\left(\frac{L}{2}-2\right)k}|w_2\rangle \right) \nonumber \\
  |\chi_b\rangle &=& \frac{1}{\sqrt{2}} \left(|w_4\rangle -e^{-i \left(\frac{L}{2}+1 \right)k }|w_3\rangle
  \right).
  \label{flatband2}
\end{eqnarray}
Note that, whereas the states in the previous example belong to
$I=+1$ sector, the basis states in Eq. \eqref{flatband2} belong to
$I=-1$ sector for $k=0,\pi$ sectors. $H_3$ has the following
representation in the basis of $|\chi_{a,b}\rangle$
\begin{eqnarray}
  \left(\begin{array}{cc} 0 & e^{ i \left(\frac{L}{2}-2 \right)k}\\ e^{ -i \left(\frac{L}{2}-2 \right)k } & 0 \end{array}\right)
  \label{flatband3}
\end{eqnarray}
and thus leads to a $2\times 2$ fragment due to a secondary fragmentation. On
diagonalization, we obtain two eigenvectors at each $k$ as follows
\begin{eqnarray}
  \frac{1}{\sqrt{2}}(\pm e^{  i \left((L/2)-2 \right)k } |\chi_a\rangle +|\chi_b\rangle)
\label{flatband4}
\end{eqnarray}
with eigenvalues $E=\pm 1$ for all $k$, thus realizing two perfectly
flat bands of anomalous eigenstates in momentum space. These flat
bands are shown in Fig.~\ref{fig4} for $L=18$ using
momentum-resolved ED results.

\section{Adding non-commuting interactions to the model}
\label{sec:otherint} We now explore the effects of adding two
non-commuting interactions to the $H_3$ model respectively (while
the constrained Hilbert space still remains the same). The first
case will be that of a staggered magnetic field while
the second will be the PXP Hamiltonian.

\subsection{Staggered magnetic field term}
\label{subsec:staggered}

The staggered magnetic field corresponds to a Hamiltonian of the
form
\begin{eqnarray}
  H_s = \sum_{j=1}^L (-1)^j \sigma_j^z.
  \label{Hstagg}
 \end{eqnarray}
 $H_s$ has some properties that are worth pointing out. Unlike $H_3$ or
$H_{\mathrm{PXP}}$, $H_s$ commutes with $Q$ (Eq.\
\ref{chiral}) and anticommutes with the spatial inversion symmetry
$I$. Furthermore, $H_s$ anticommutes with the operator $C$ defined
as
\begin{eqnarray}
  C = \prod_{j=1}^{L/2}\left(\sigma^z_{2j-1}+\sigma^z_{2j}\right)
  \label{CforHs}
\end{eqnarray}
which ensures that every eigenstate $|E\rangle$ with energy $E$ has a partner
$C|E\rangle$ with energy $-E$. Together with $\{H,QI\}=0$, this implies the
presence of an exponentially large number of zero modes in the spectrum of
$H_s$. These zero modes are simply Fock states in the constrained Hilbert space
with the same number of $\uparrow$ spins (and hence $\downarrow$ spins) on the even and
odd sites of the chain.

The model that we consider now has the following Hamiltonian
\begin{eqnarray}
  H_\Delta&=&H_3+\frac{\Delta}{2}H_s  \label{HDelta} \\
  &=&\sum_{j=1}^L (\tilde{\sigma}_j^+
  \tilde{\sigma}_{j-1}^- \tilde{\sigma}_{j+1}^-
  + \mathrm{H.c.})+\frac{\Delta}{2}\sum_{j=1}^L (-1)^j \sigma_j^z \nonumber
\end{eqnarray}
It is immediately clear that the connectivity graphs of $H_3$ (Fig.~\ref{fig1})
 and $H_\Delta$
are identical in the basis of the Fock states since $H_s$ is a
purely diagonal term in this basis. Thus, $H_\Delta$ continues to
show Hilbert space fragmentation in spite of the fact that $H_s$
does not commute with $H_3$. Like $H_3$ and $H_s$, the spectrum of
$H_\Delta$ also has an $E \rightarrow -E$ symmetry. Furthermore,
since $H_\Delta$ anticommutes with the combination $QI$, it also
harbors an exponentially large number of mid-spectrum zero modes.

We first calculate the degeneracy of zero modes of $H_\Delta$ numerically using
ED for chain lengths $L=16,18,20$ when $\Delta=0$, $\Delta \neq 0$, and
$\Delta \rightarrow \infty$ (non-interacting limit) and give the numbers in
Table.~\ref{table2}.
We also give a lower bound for the number of zero modes at $\Delta \neq 0$
in the same table which we will justify below.
Interestingly, the numerics indicate that the number of zero modes is independent of $\Delta$ as long as it is finite and non-zero.
\begin{table}
  \caption{The number of zero modes obtained from exact diagonalization shown
    for three chain lengths $L=16,18,20$ for $\Delta=0$, $\Delta \neq 0$, and
    $\Delta \rightarrow \infty$. A lower bound is also given for the number of
  zero modes for $\Delta \neq 0$.}
  \label{table2}
\begin{center}
  \begin{tabular}{|c|c|c|c|c|}
    \hline \hline
    $L$ & $\Delta=0$ & $\Delta \neq 0$ & Lower bound &$\Delta \rightarrow \infty$  \\
    \hline \hline
    $16$ & $193$ & $77$ & $65$ & $313$ \\
    \hline
     $18$ & $348$ & $180$ & $91$ & $778$\\
    \hline
    $20$ & $709$ & $247$ & $151$  & $1941$ \\
    \hline
    \end{tabular}
\end{center}
\end{table}

We first consider bubble type fragments with $n$ bubbles.
  The case of $n=0$ is the simplest. The inert state of $H_3$
continues to be the lone inert state of $H_\Delta$ and is a
simultaneous eigenket of both the non-commuting terms for any $L$.
We now follow the same logic used to derive Eq.~\ref{bubbleHfree}
for $n \neq 0$ and again denote $X_{1,j}$ and $X_{2,j}$ as
pseudospin-$1/2$ variables with $\tau_j^z=+1$ and $\tau_j^z=-1$.
Here, it is useful to note
that the energy of a bubble Fock state with respect to $H_s$
can be written in the form $E_s=\sum_j(E_j(X_{1,j})+E_j(X_{2,j}))$
where $E_j(X_{1,j})$ equals $-4 (4)$ if $j$ is even (odd) and
$E_j(X_{2,j})$ equals $2 (-2)$ if  $j$ is even (odd) with $X_{1,j}
(X_{2,j})$ representing a $\uparrow_{j-1}\downarrow_j \uparrow_{j+1}
(\downarrow_{j-1}\uparrow_j \downarrow_{j+1})$ unit of three
consecutive spins as defined in Eq.~\ref{bubstates}.
Then, the form of $H_\Delta$ projected to any $n \neq 0$ bubble type fragment
simply equals
\begin{widetext}
\begin{eqnarray}
  H_{\Delta,\mathrm{bubble}} &=& \sum_{j_b=1}^{n} \tau_{j_b}^x +\frac{3\Delta}{2}
  (\sum_{j_b=1}^n (-1)^{j_b}\tau_{j_b}^z)+\frac{\Delta}{2}(n_{\mathrm{odd}}-n_{\mathrm{even}}) \nonumber \\
  &=& \frac{1}{2} \sqrt{4+9\Delta^2} \left( \sum_{j_b=1}^{n} \hat{h} \cdot \tilde{\mathbf{\tau}}_{j_b} \right)+\frac{\Delta}{2}(n_{\mathrm{odd}}-n_{\mathrm{even}})
  \label{bubbleHDfree}
\end{eqnarray}
\end{widetext}
where $j_b$ denotes the central sites of the bubbles with
$n_{\mathrm{odd}}$ ($n_{\mathrm{even}}$) equalling the number of
bubbles centered on odd (even) sites of the lattice. The second line
of Eq.~\ref{bubbleHDfree} has been obtained from the first line by
using the transformation $\tilde{\tau}_j^{y,z} =-\tau_j^{y,z}$ for
any $j$ that is odd. The unit vector $\hat{h}$ equals
$(2/\sqrt{4+9\Delta^2},0,3\Delta/\sqrt{4+9\Delta^2})$. Since this
``non-interacting'' effective Hamiltonian cannot entangle the
$\tilde{\tau}_j$ variables, all the eigenstates of $H_\Delta$
obtained from the bubble fragments continue to have a strict area
law scaling of entanglement entropy even when $L \gg 1$.

Comparing Eq.~\ref{bubbleHDfree} with Eq.~\ref{bubbleHfree}, we
immediately see that for any fragment with an equal number of
bubbles centered on even and odd sites of the lattice, the
eigenvalues of $H_\Delta$ and $H_3$ are just related by a scale
factor that equals $(\sqrt{4+9\Delta^2})/2$. In particular, the
degeneracy of the zero modes obtained from such fragments remain
unchanged with $\Delta$ even though the eigenvectors themselves
change. In contrast, the degeneracy of zero modes drops to zero for
bubble type fragments where the $n$ bubbles are distributed
unequally between even and odd sites of the chain when $\Delta$ is
changed from zero to any non-zero value because of the
$(\Delta/2)(n_{\mathrm{odd}}-n_{\mathrm{even}})$ term in
Eq.~\ref{bubbleHDfree}. We have used these two facts to obtain a
lower bound on the number of zero modes in $H_\Delta$ in
Table.~\ref{table2}.

We now explicitly look at the cases with $n=2$ and $n=4$ from
which we see that not only does the number of zero modes for such
fragments stay independent of $\Delta$ but the zero modes come in
two varieties -- those that are simultaneous eigenkets of $H_s$ and
$H_3$ and those that are not. The eigenfunctions of the former class
of zero modes stay unchanged with $\Delta$ while the latter class of
zero modes change as $\Delta$ is varied.
Below, we will assume that half of these units are
centered on even sites while the other half are centered on odd
sites of the chain.

{{\it{$n=2$ case}}: Consider the following four bubble-type Fock
states
  \begin{eqnarray}
    |b_1\rangle &=& \cdots \downarrow \downarrow \boxed{X_{2,i}} \downarrow \cdots \downarrow \boxed{X_{1,j}} \downarrow \downarrow \cdots (E_s=+6) \nonumber \\
    |b_2\rangle &=& \cdots \downarrow \downarrow \boxed{X_{2,i}} \downarrow \cdots \downarrow \boxed{X_{2,j}} \downarrow \downarrow \cdots (E_s=0) \nonumber \\
    |b_3\rangle &=& \cdots \downarrow \downarrow \boxed{X_{1,i}} \downarrow \cdots \downarrow \boxed{X_{1,j}} \downarrow \downarrow \cdots (E_s=0) \nonumber \\
    |b_4\rangle &=& \cdots \downarrow \downarrow \boxed{X_{1,i}} \downarrow \cdots \downarrow \boxed{X_{2,j}} \downarrow \downarrow \cdots (E_s=-6) \nonumber \\
\label{ne2}
  \end{eqnarray}
where we choose $i$ to be an even site and $j$ an odd site on the
chain. These form a $4 \times 4$ fragment for $H_\Delta$ at any
$\Delta$ and are eigenstates of $H_\Delta$ when $\Delta \rightarrow
\infty$. When $\Delta=0$, the eigenstates are instead bubble-type
eigenstates of $H_3$ (as discussed Sec.~\ref{sec:bubblestates}) with
energies $+2$ (with degeneracy $1$), $0$ (with degeneracy $2$) and
$-2$ (with degeneracy $1$).
The (unnormalized) eigenvectors for the two zero modes are as
follows:
  \begin{eqnarray}
    |z_1\rangle &=& |b_2\rangle -|b_3\rangle \nonumber \\
    |z_2 \rangle &=& (|b_1\rangle -|b_4\rangle)-3\Delta |b_2\rangle.
    \label{ne2zeromodes}
  \end{eqnarray}
While $|z_1 \rangle$ is composed of the zero modes of $H_s$ and is,
therefore, a simultaneous eigenket of $H_3$ and $H_s$, $|z_2\rangle$
is not an eigenket of $H_3$ or $H_s$ but only of $H_\Delta$.

  {{\it{$n=4$ case}}: Consider the following sixteen bubble-type Fock
  states
    \begin{widetext}
  \begin{eqnarray}
    |c_1\rangle &=& \cdots \downarrow \downarrow \boxed{X_{2,i}} \downarrow \cdots \downarrow \boxed{X_{2,j}} \downarrow \cdots \downarrow \boxed{X_{1,l}} \downarrow \cdots \downarrow \boxed{X_{1,m}}\downarrow \downarrow \cdots (E_s=+12) \nonumber \\
    |c_2\rangle &=& \cdots \downarrow \downarrow \boxed{X_{2,i}} \downarrow \cdots \downarrow \boxed{X_{2,j}} \downarrow \cdots \downarrow \boxed{X_{1,l}} \downarrow \cdots \downarrow \boxed{X_{2,m}}\downarrow \downarrow \cdots (E_s=+6) \nonumber \\
    |c_3\rangle &=& \cdots \downarrow \downarrow \boxed{X_{2,i}} \downarrow \cdots \downarrow \boxed{X_{2,j}} \downarrow \cdots \downarrow \boxed{X_{2,l}} \downarrow \cdots \downarrow \boxed{X_{1,m}}\downarrow \downarrow \cdots (E_s=+6) \nonumber \\
    |c_4\rangle &=& \cdots \downarrow \downarrow \boxed{X_{1,i}} \downarrow \cdots \downarrow \boxed{X_{2,j}} \downarrow \cdots \downarrow \boxed{X_{1,l}} \downarrow \cdots \downarrow \boxed{X_{1,m}}\downarrow \downarrow \cdots (E_s=+6) \nonumber \\
    |c_5\rangle &=& \cdots \downarrow \downarrow \boxed{X_{2,i}} \downarrow \cdots \downarrow \boxed{X_{1,j}} \downarrow \cdots \downarrow \boxed{X_{1,l}} \downarrow \cdots \downarrow \boxed{X_{1,m}}\downarrow \downarrow \cdots (E_s=+6) \nonumber \\
    |c_6\rangle &=& \cdots \downarrow \downarrow \boxed{X_{1,i}} \downarrow \cdots \downarrow \boxed{X_{1,j}} \downarrow \cdots \downarrow \boxed{X_{1,l}} \downarrow \cdots \downarrow \boxed{X_{1,m}}\downarrow \downarrow \cdots (E_s=0) \nonumber \\
    |c_7\rangle &=& \cdots \downarrow \downarrow \boxed{X_{1,i}} \downarrow \cdots \downarrow \boxed{X_{2,j}} \downarrow \cdots \downarrow \boxed{X_{1,l}} \downarrow \cdots \downarrow \boxed{X_{2,m}}\downarrow \downarrow \cdots (E_s=0) \nonumber \\
    |c_8\rangle &=& \cdots \downarrow \downarrow \boxed{X_{1,i}} \downarrow \cdots \downarrow \boxed{X_{2,j}} \downarrow \cdots \downarrow \boxed{X_{2,l}} \downarrow \cdots \downarrow \boxed{X_{1,m}}\downarrow \downarrow \cdots (E_s=0) \nonumber \\
    |c_9\rangle &=& \cdots \downarrow \downarrow \boxed{X_{2,i}} \downarrow \cdots \downarrow \boxed{X_{1,j}} \downarrow \cdots \downarrow \boxed{X_{2,l}} \downarrow \cdots \downarrow \boxed{X_{1,m}}\downarrow \downarrow \cdots (E_s=0) \nonumber \\
    |c_{10}\rangle &=& \cdots \downarrow \downarrow \boxed{X_{2,i}} \downarrow \cdots \downarrow \boxed{X_{1,j}} \downarrow \cdots \downarrow \boxed{X_{1,l}} \downarrow \cdots \downarrow \boxed{X_{2,m}}\downarrow \downarrow \cdots (E_s=0) \nonumber \\
    |c_{11}\rangle &=& \cdots \downarrow \downarrow \boxed{X_{2,i}} \downarrow \cdots \downarrow \boxed{X_{2,j}} \downarrow \cdots \downarrow \boxed{X_{2,l}} \downarrow \cdots \downarrow \boxed{X_{2,m}}\downarrow \downarrow \cdots (E_s=0) \nonumber \\
    |c_{12}\rangle &=& \cdots \downarrow \downarrow \boxed{X_{1,i}} \downarrow \cdots \downarrow \boxed{X_{1,j}} \downarrow \cdots \downarrow \boxed{X_{1,l}} \downarrow \cdots \downarrow \boxed{X_{2,m}}\downarrow \downarrow \cdots (E_s=-6) \nonumber \\
    |c_{13}\rangle &=& \cdots \downarrow \downarrow \boxed{X_{1,i}} \downarrow \cdots \downarrow \boxed{X_{1,j}} \downarrow \cdots \downarrow \boxed{X_{2,l}} \downarrow \cdots \downarrow \boxed{X_{1,m}}\downarrow \downarrow \cdots (E_s=-6) \nonumber \\
    |c_{14}\rangle &=& \cdots \downarrow \downarrow \boxed{X_{1,i}} \downarrow \cdots \downarrow \boxed{X_{2,j}} \downarrow \cdots \downarrow \boxed{X_{2,l}} \downarrow \cdots \downarrow \boxed{X_{2,m}}\downarrow \downarrow \cdots (E_s=-6) \nonumber \\
    |c_{15}\rangle &=& \cdots \downarrow \downarrow \boxed{X_{2,i}} \downarrow \cdots \downarrow \boxed{X_{1,j}} \downarrow \cdots \downarrow \boxed{X_{2,l}} \downarrow \cdots \downarrow \boxed{X_{2,m}}\downarrow \downarrow \cdots (E_s=-6) \nonumber \\
    |c_{16}\rangle &=& \cdots \downarrow \downarrow \boxed{X_{1,i}} \downarrow \cdots \downarrow \boxed{X_{1,j}} \downarrow \cdots \downarrow \boxed{X_{2,l}} \downarrow \cdots \downarrow \boxed{X_{2,m}}\downarrow \downarrow \cdots (E_s=-12)
    \label{ne4}
  \end{eqnarray}
  \end{widetext}
where we choose $i$, $j$ to be an even sites and $l$, $m$ to be odd
sites on the chain. These Fock states form a $16 \times 16$ fragment
for $H_\Delta$ and
    are eigenstates of $H_\Delta$ when $\Delta \rightarrow \infty$. When $\Delta=0$, the eigenstates of $H_\Delta$ are instead bubble-type eigenstates (as discussed Sec.~\ref{sec:bubblestates}) with energies $+4$ (with degeneracy $1$), $+2$ (with
    degeneracy $4$), $0$ (with degeneracy $6$), $-2$ (with degeneracy $4$) and
    $-4$ (with degeneracy $1$).
    Writing
    the (unnormalized)
    eigenvectors for the  zero modes:
    \begin{widetext}
      \begin{eqnarray}
        |z_1\rangle &=& |c_6\rangle -|c_7\rangle -|c_9\rangle +|c_{11}\rangle \nonumber \\
        |z_2\rangle &=& |c_7\rangle -|c_8\rangle +|c_9\rangle -|c_{10}\rangle \nonumber \\
        |z_3\rangle &=& (|c_3\rangle -|c_4\rangle +|c_{12}\rangle -|c_{15}\rangle)+3\Delta(|c_6\rangle -|c_9\rangle) \nonumber \\
        |z_4\rangle &=& (|c_3\rangle -|c_5\rangle +|c_{12}\rangle -|c_{14}\rangle)+3\Delta(|c_6\rangle -|c_8\rangle) \nonumber \\
        |z_5\rangle &=& (|c_2\rangle -|c_3\rangle -|c_{12}\rangle +|c_{13}\rangle)-3\Delta(|c_7\rangle -|c_8\rangle) \nonumber \\
        |z_6\rangle &=& (|c_1\rangle -|c_{16}\rangle) -3\Delta(|c_2\rangle -|c_3\rangle +|c_{4}\rangle -|c_{5}\rangle -2|c_{12}\rangle)+[9\Delta^2(|c_6\rangle +|c_7\rangle -|c_8\rangle)-(|c_7\rangle+|c_9\rangle)]
        \label{ne4zeromodes}
      \end{eqnarray}
    \end{widetext}
We also see a similar structure in the eigenvectors of the zero
modes when compared to the $n=2$ case. While $|z_{1,2}\rangle$ are
simultaneous eigenkets of both $H_3$ and $H_s$ since these are
entirely composed of the zero modes of $H_s$, this is not the case
for the rest of the zero modes which keep changing as a function of
$\Delta$.

For a general $n$, the total number of such zero modes in a given
bubble fragment} can be estimated as follows. First, we note from
the previous discussion that the number of such zero energy states
are same as the number of zero energy eigenstates of $H_s$. Thus we
need to estimate the number of such zero energy states for arbitrary
$n$. To this end, we note that for $E_s=0$, we need to have equal
number of $X_{1,j}$ and $X_{2,j}$ bubbles on even and odd sites. Let
us then consider a configuration where there are $p$ $X_{1,j}$
bubbles on even sublattice. This implies that one must have $(n/2 -
p)$ $X_{2,j}$ bubbles on even sublattice. Also such a configuration
can be realized in $^{n/2}C_{p}\,\, ^{n/2}C_{n/2-p}$ ways. Since $p$
can range from $0$ to $n/2$. we find the total number of zero energy
states for $n$ bubble states to be
\begin{eqnarray}
N_0 &=& \sum_{p=0}^{n/2} \left( ^{n/2}C_{p} \right)^2=\left(^{n}C_{n/2} \right)
\label{zeronum}
\end{eqnarray}
where we have used the relation $^{n/2}C_{p}= ^{n/2}C_{n/2-p}$. Eq.\
\ref{zeronum} correctly estimate $N_0=2(6)$ for $n=2(4)$ as obtained
earlier. We note that this counting assumes $L \gg n$ and does not
estimate the number of ways $n$ can be realized on a chain of length
$L$. Further, the equality between zero energy of eigenstates and
$H_s$ and $H_{\Delta}$ only pertains to their number; the states, as
shown above in Eqs.\ \ref{ne2zeromodes} and \ref{ne4zeromodes}, are
indeed distinct. The number of simultaneous zero modes of $H_3$ and
$H_s$ for $n=0,2,\cdots,12$ is given in Table.~\ref{table3} for
comparison.
\begin{table}
  \caption{The simultaneous zero modes of $H_s$ and $H_3$ and the total
    number of zero modes of $H_\Delta$ shown for a fragment with $n$ bubbles
  where equal number of bubbles occupy even and odd sites.}
  \label{table3}
\begin{center}
  \begin{tabular}{|c|c|c|}
    \hline \hline
    $n$ & Simultaneous zero modes & Total number of zero modes \\
    \hline \hline
    $0$ & $1$ & $1$ \\
    \hline
    $2$ & $1$ & $2$ \\
    \hline
    $4$ & $2$ & $6$ \\
    \hline
    $6$ & $5$ & $20$ \\
    \hline
    $8$ & $14$ & $70$ \\
    \hline
    $10$ & $42$ & $252$ \\
    \hline
    $12$ & $132$ & $924$ \\
    \hline
    \end{tabular}
\end{center}
\end{table}

Just like the bubble-type fragments,
  the fate of the primary fragments in momentum space discussed in
  Sec.~\ref{sec:nonbubblestates} can also be understood for non-zero $\Delta$.
  The $2 \times 2$ primary fragments defined for each of the $L$ allowed momenta
  $k$ at $\Delta=0$ now get converted to $4 \times 4$ primary fragments
  defined at $L/2$ momenta $k=2\pi n/L$ where $n=0,1,2,\cdots,\frac{L}{2}-1$ at
  any finite non-zero $\Delta$
  since the $H_s$ term connects states with momentum $k$ and $k+\pi$. Writing
  the matrix representation of $H_\Delta$ in the primary fragment represented by
  $|\Phi_{1,k}\rangle, |\Phi_{2,k}\rangle, |\Phi_{1,k+\pi}\rangle$ and
  $|\Phi_{2,k+\pi}\rangle$ (Eq.~\ref{fragmentink}), we get the following
matrix
\begin{eqnarray}
  \left(\begin{array}{cccc} 0 & 1+e^{ -ik}& 0 & 0\\ 1+e^{ ik} & 0 &0 &3\Delta \\
  0 & 0 &0 & 1+e^{ -i(k+\pi)}\\ 0 & 3\Delta & 1+e^{ i(k+\pi)} & 0 \end{array}\right)
  \label{4times4}
\end{eqnarray}
which gives four anomalous eigenstates at each $k$ (with $k$ and $k+\pi$
identified) with the eigenvalues
\begin{eqnarray}
E(k) = \pm \frac{\sqrt{\mathcal{A}(\Delta) \pm \sqrt{\mathcal{A}(\Delta)^2-16 \sin (k)^2}}}{\sqrt{2}}
\end{eqnarray}
where $\mathcal{A}(\Delta) = 4+9\Delta^2$. For $k=0$, this immediately gives
two zero modes for $H_\Delta$ with one of them being $|\Phi_{1,k=\pi} \rangle$
and the other being $-\frac{3\Delta}{2}|\Phi_{1,k=0} \rangle+|\Phi_{2,k=\pi} \rangle$. While the first zero mode is a simultaneous eigenket of both $H_3$ and
$H_s$, the second zero mode is not and changes as a function of $\Delta$.
Table.~\ref{table2} clearly shows that
there are several other zero modes that survive in the non-bubble type
fragments when a finite $\Delta$ is turned on. We do not have a
complete understanding of this phenomenon.

\subsection{Floquet version}
\label{subsec:Floquet}

The form of $H_\Delta$ (Eq.~\ref{HDelta})
  allows us to naturally
  define an interacting Floquet problem with a time-dependent
  $\Delta(t)$ that again shows Hilbert space fragmentation.
  The simplest setting for such a drive, which we shall consider here,
  constitutes a square pulse protocol for which
\begin{eqnarray}
\Delta(t) &=&  -\Delta_0 \quad  t \le T_0/2 \nonumber\\
&=& \Delta_0 \quad  t> T_0/2
\end{eqnarray}
where $T_0 = 2 \pi/\omega_D$ is the time period and $\omega_D$ is
the drive frequency. While the spectrum of the single period
evolution operator $U(T_0,0)$ resembles that of $H_3$ when $\omega_D \gg
1,\Delta_0$
(the high-drive frequency limit), the situation is markedly different in
the regime with moderate or low drive frequency since the Floquet Hamiltonian
$H_F$ defined using $U(T_0,0)=\exp(-iH_FT_0)$ contains longer-ranged terms in
real space. This can be seen from Fig.~\ref{floquet} where we show the
half-chain entanglement entropy $S_{L/2}$ for the eigenstates of $U(T,0)$ with
$\Delta_0=\omega_D=1$ from ED using $L=24$ and $k=0$ (mod $\pi$).
While the majority of eigenstates have $S_{L/2}$ close
to $S_{\mathrm{Page}}$ as expected of interacting Floquet systems, it is clear
from the numerical data 
that there are also several anomalous eigenstates with much lower entanglement.
We explain a class of these anomalous eigenstates in $U(T,0)$ below.

First, we note a few consequences of the simple local form of
Eq.\ \ref{bubbleHDfree} when the system is driven periodically from
any initial state which is a bubble-type Fock state.
In what follows, we shall also consider
$n_e=n_0$ without any loss of generality. Thus the driven
Hamiltonian in the bubble sector can be written as
\begin{eqnarray}
H_{\Delta,{\rm bubble}}(t) &=& \sum_{j_b} \left(\tau_{j_b}^x  +
\frac{3 \Delta(t)}{2} \tau_{j_b}^z (-1)^{j_b}\right)
\label{timeHbubble}
\end{eqnarray}
We note that the driven Hamiltonian is analogous to that for a
collection of non-interacting spin-half particles. This allows to
solve for the evolution operator $U(T_0,0) =
\prod_{j=1}^{L} U_j(T_0,0)$  exactly; we find
\begin{eqnarray}
U_j(T_0,0) &=& \left( \begin{array}{cc} p & q_{j_b} \\
-q_{j_b}^{\ast} & p
\end{array} \right), \quad
p = \frac{ (3\Delta_0/2)^2 +\cos \epsilon_0 T_0}{\epsilon_0^2}
\nonumber\\
q_{j_b} &=& -\frac{ \left\{1-\cos(\epsilon_0T_0)\right\}(-1)^{j_b}\left(\frac{3\Delta_0}{2}\right) + i \epsilon_0 \sin (\epsilon_0 T_0)}{\epsilon_0^2}
\nonumber\\ \label{evolbub}
\end{eqnarray}
where $\epsilon_0= \sqrt{1+ (3 \Delta_0/2)^2}$. Thus we find that
$U(T_0,0)=I$ for $\omega_D=  \omega_{\rm f}= \epsilon_0/n$ for
integer $n$. This shows that at these drive frequencies, the
stroboscopic dynamics of any state in the bubble sector will be
frozen for $\omega_D= \omega_{\rm f}$. Moreover, the presence of
such an on-site structure of the evolution operator $U(T_0,0)$
ensures that the corresponding Floquet eigenstates will not
thermalize at any drive frequency; this provides yet another
manifestation of ETH violation in such models where an exponentially
large number of Floquet eigenstates satisfy area law scaling of
entanglement entropy even for $L \gg 1$. We note that
in contrast to earlier works \cite{adas1,dpekker1,dyn4}, the model
displays freezing for infinite number of drive frequencies and an
exponentially large number of initial states.

Similarly, generalizing Eq.~\ref{4times4} to the case with a
time-periodic $\Delta(t)$ for the primary fragments in momentum
space and exponentiating the $4 \times 4$ matrices immediately shows
that the Floquet unitary $U(T_0,0)$ has an emergent $SU(4)$
structure for such non-bubble fragments even though the basic
degrees are $S=1/2$ spins. These primary fragments will again
generate ETH-violating Floquet eigenstates at any drive frequency.
E.g., the state $|\Phi_{1,k=\pi} \rangle$ is a zero mode of $U(T,0)$
for any $\omega_D$.

In contrast, we expect that the secondary fragmentation discussed for $H_3$
will not survive in this Floquet version, especially for small drive
frequencies $\omega_D$ and the other primary fragments (apart than the cases
discussed above) will show thermalization for the corresponding Floquet
eigenstates. The details of such
phenomena and the behavior of the model for other drive protocols is
left as subject of future work. 

\begin{figure}[!htb]
  \includegraphics[width=0.96\hsize]{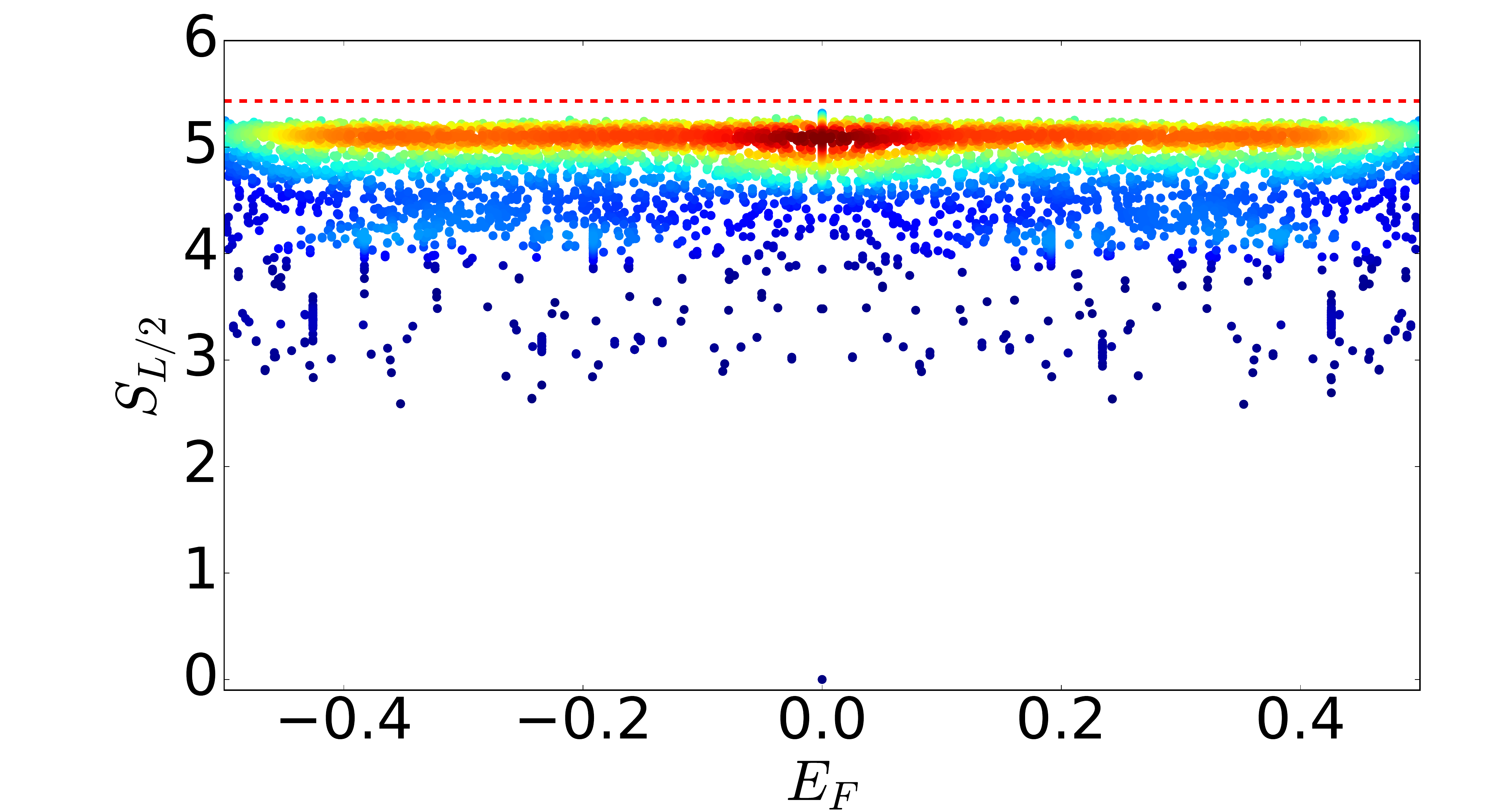}
  \caption{The bipartite entanglement entropy with equal partitions, $S_{L/2}$,
    shown for all the eigenstates of the Floquet unitary $U(T,0)$ with
    $\Delta_0=\omega_D=1$ for $L=24$ and $k=0$ (mod $\pi$). $E_F$ denotes the
    eigenvalue of $H_F$ and is defined in the interval $[-\pi/T,\pi/T)$.
      The horizontal
    dotted line indicates the average entanglement entropy, $S_{\mathrm{Page}}$,
    of random pure states. The density of states is indicated by a color map
  with warmer colors indicating higher density of states.}
  \label{floquet}
\end{figure}

\subsection{PXP term}
\label{subsec:PXP}

In this subsection, we consider the effect of adding a PXP
Hamiltonian to $H_3$ so that
\begin{eqnarray}
H_{\alpha} &=& H_3 + \alpha H_{\rm PXP}  \label{3pxpham}
\end{eqnarray}
where $H_3$ and $H_{\rm PXP}$ are given by Eqs.\ \ref{H3def} and
\ref{Hpxpdef} respectively. Here $\alpha$ is a dimensionless
parameter which determine the relative strength of the two terms in
Eq.\ \ref{3pxpham}; in what follows, we shall restrict ourselves to
$\alpha \ll 1$ and explore the effect of $H_{\rm PXP}$ in this
perturbative regime starting from bubble initial states.

To understand the effect of $H_{\rm PXP}$ within the bubble
manifold, we note $H_{\rm PXP}$ provides a dispersion to these
bubbles which are strictly localized under the action of $H_3$. To
justify this statement, let us first consider the effect of $H_{\rm
PXP}$ on a bubble Fock state $|X_{1,j}\rangle$. It can be checked
that
\begin{eqnarray}
\sum_{j_1} \tilde \sigma^x_{j_1} |X_{1,j}\rangle &=& |\Phi_{2,j-1}
\rangle + |\Phi_{2,j-3}\rangle \nonumber\\
&& + |X_{2,j-1}\rangle + |X_{2,j+1}\rangle + |\psi'\rangle
\nonumber\\
|\psi'\rangle &=& |X_{1,j}, X_{2,\ell}\rangle
 \label{x1eqj}
\end{eqnarray}
where $\ell \ne j, j\pm 1, j\pm 2, j\pm 3$, $|\psi'\rangle$
represent a state with one $X_1$ and one $X_2$ bubble whose centers
are separated by at least three sites, and $|\Phi_{1(2) j}\rangle$
are given by Fourier transform of states defined Eq.\
\ref{fragmentink}. Note that $|\psi'\rangle$ takes the system out of
the subspace spanned by states $|X_{1,j}\rangle$, $|X_{2,j}\rangle$,
$|\Phi_{1,j}\rangle$, and $|\Phi_{2,j}\rangle$. In what follows, we
are going to ignore all such states. This step will be justified a
posteriori while studying quantum dynamics induced by $H_{\alpha}$
starting from an initial state $|X_{1,j}\rangle$ in
Sec.~\ref{sec:quenches}. Here we note that the effect of
disregarding these additional state can be thought to lead to a
decay rate, as per Fermi golden rule, leading to a loss of weight of
the state $|X_{1,j}\rangle$ within the subspace. Such a decay rate
clearly scales as $\alpha^2/\Delta E$ where $\Delta E$ denotes the
energy difference of the eigenstates outside the subspace which have
a finite matrix element with eigenstates within it. We have not been
able to estimate $\Delta E$ in this work; we merely note that as
long as $\Delta E \sim \mathrm{O}(\alpha)$ or larger, the decay rate
will be small for small $\alpha$. We are going to address this issue
while discussing dynamics induced by $H_{\alpha}$.

Ignoring the states $|\psi'\rangle$, and noting $H_3 |X_{1,j}\rangle
= |X_{2,j}\rangle$, we find that a Fourier transform of Eq.\
\ref{x1eqj} allows us to write
\begin{eqnarray}
H_{\alpha} |X_{1,k}\rangle &=& \epsilon_k  |X_{2,k}\rangle + c_{2k}
|\Phi_{2,k} \rangle \label{h2eq1} \\
\epsilon_{k} &=& (1+2 \alpha \cos k), \quad c_{2k}= \alpha( e^{-ik}
+ e^{-3 ik} ),  \nonumber
\end{eqnarray}
where $|X_{1,k}\rangle = \sum_j \exp[i k j]
|X_{1,j}\rangle$. An exactly similar consideration shows (ignoring
the states outside the subspace)
\begin{eqnarray}
H_{\alpha} |X_{2,k}\rangle &=& \epsilon_k |X_{1,k}\rangle +c_{1k}
|\Phi_{1,k}\rangle \nonumber\\
H_{\alpha} |\Phi_{1,k} \rangle &=& \beta_k |\Phi_{2,k}\rangle +
c_{1k}^{\ast}
|X_{2,k}\rangle, \nonumber\\
H_{\alpha} |\Phi_{2,k}\rangle &=&  \beta_k^{\ast} |\Phi_{1,k}\rangle +
c_{2k}^{\ast}
|X_{1,k}\rangle \nonumber\\
\beta_k &=& (1+ e^{ik}), \quad c_{1k}= \alpha(1+ e^{-3ik})
\label{h2eq2}
\end{eqnarray}
Thus finding the eigenvalues of $H_{\alpha}$ in this subspace
amounts to  diagonalization of the $4 \times 4$ matrix given by
\begin{eqnarray}
\left (\begin{array}{cccc} 0 & \epsilon_k & c_{2k} & 0 \\
\epsilon_k & 0 & 0 & c_{1k} \\
c_{2k}^{\ast}& 0 & 0 & \beta_k \\
0 & c_{1k}^{\ast} & \beta_k^{\ast} & 0 \end{array} \right)
\end{eqnarray}
The four eigenvalues $E_{i k}$ for $i=1,2,3,4$ obtained by
diagonalization of this matrix is shown in Fig.\ \ref{figpert} We
find that for $\alpha=0$, the action of $H_3$ correctly reproduces
the flat band (corresponding to the subspace spanned by
$|X_{1,k}\rangle$ and $|X_{2,k}\rangle$) and the dispersing band with
$E_K=2 \cos k/2$ (corresponding to the subspace spanned by
$|\Phi_{1,k}\rangle$ and $|\Phi_{2,k}\rangle$). The states within each
of these $2 \times 2$ subspaces do not interact with those of the
other subspace for $\alpha=0$ as seen from the top left panel of
Fig.\ \ref{figpert}. Note the bands touch at $k=\pm 2\pi/3$.
\begin{figure}[!tbp]
\includegraphics[width=0.49\hsize]{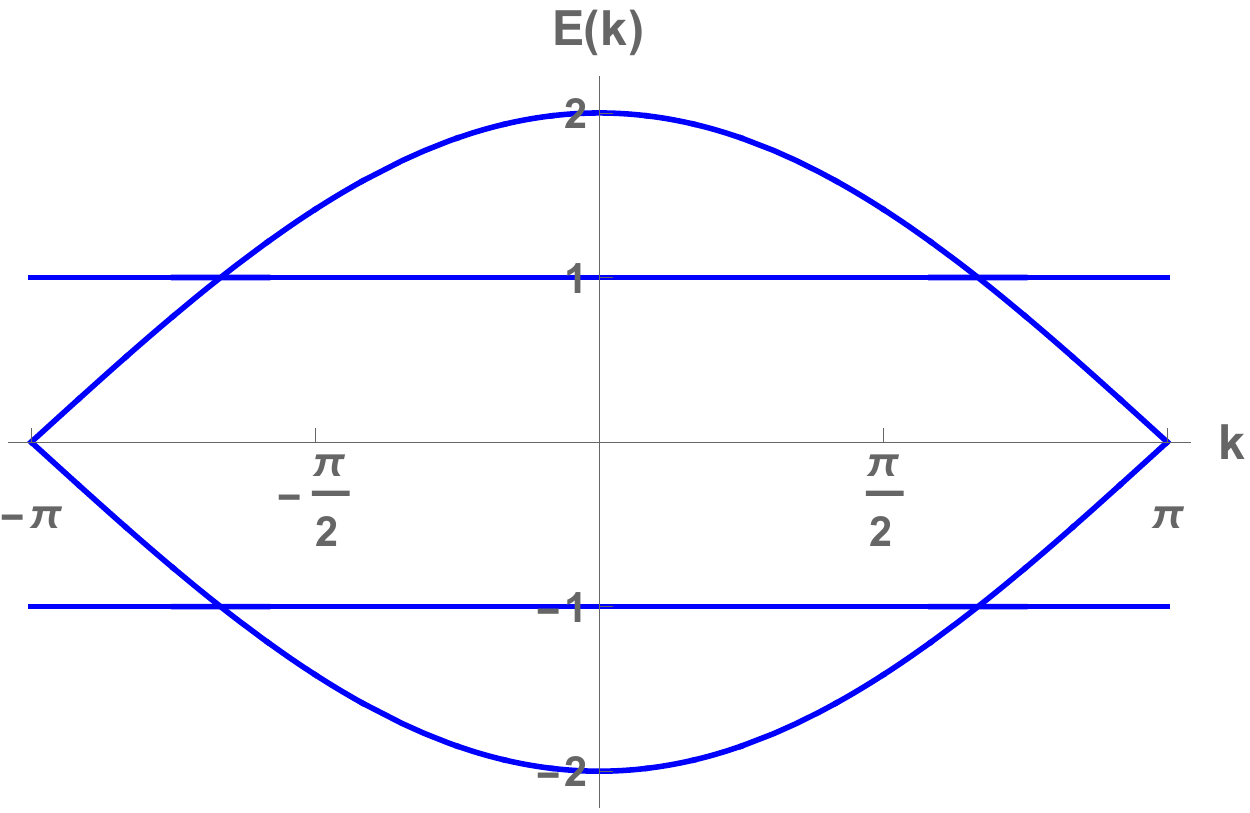}
\includegraphics[width=0.49\hsize]{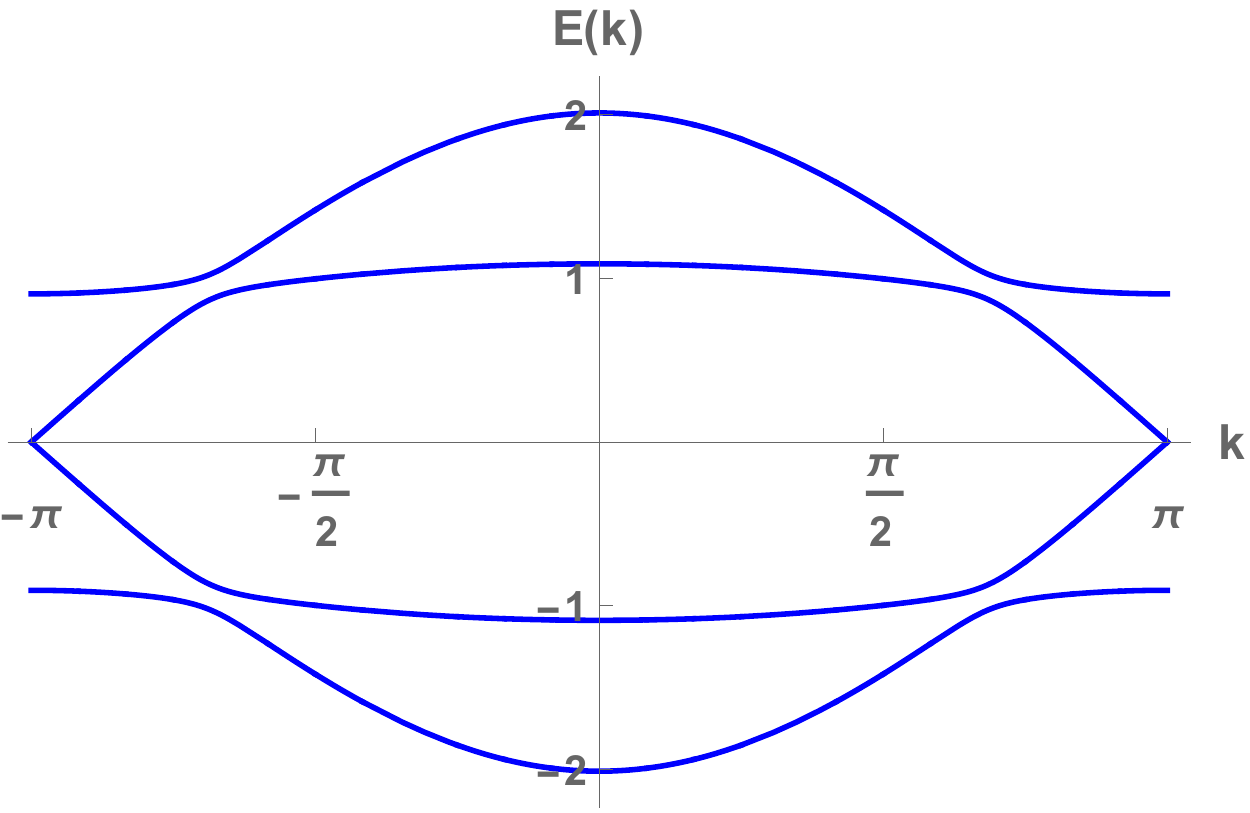}
\includegraphics[width=0.49\hsize]{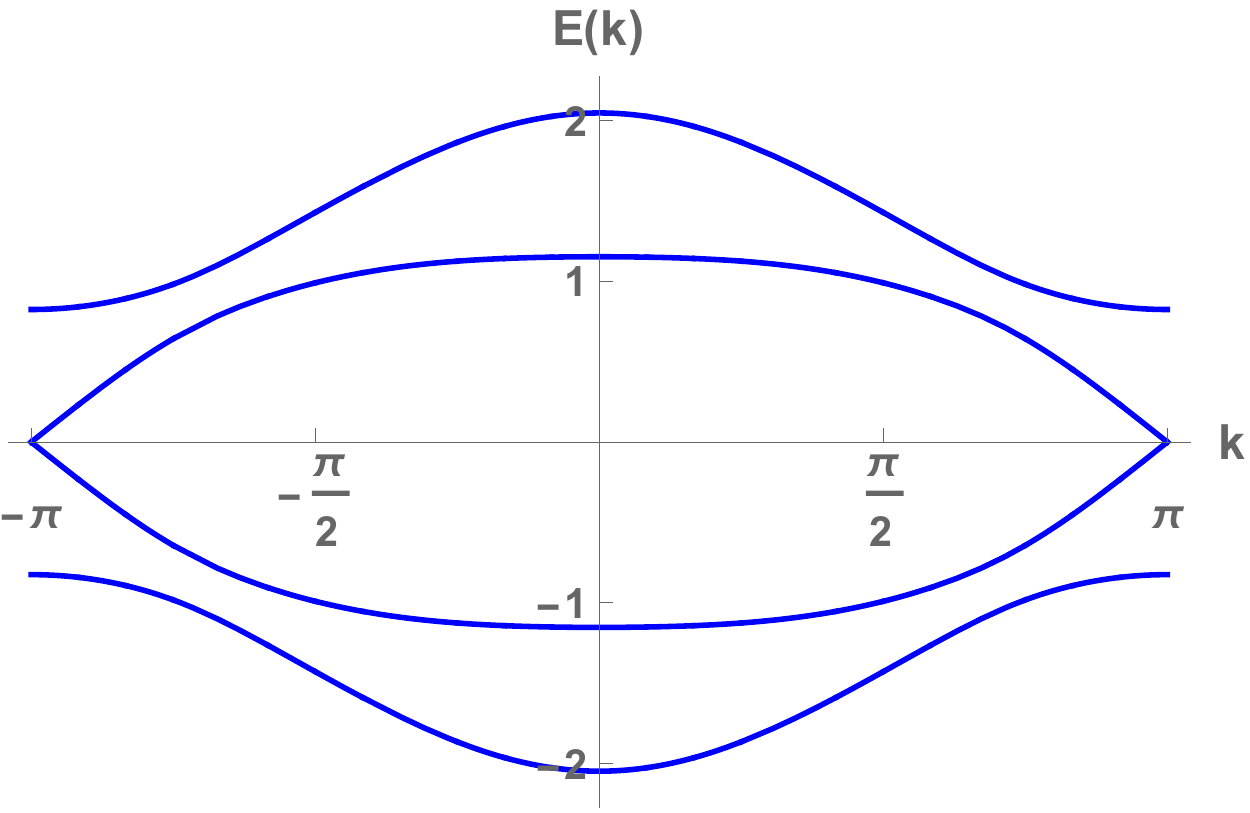}
\includegraphics[width=0.49\hsize]{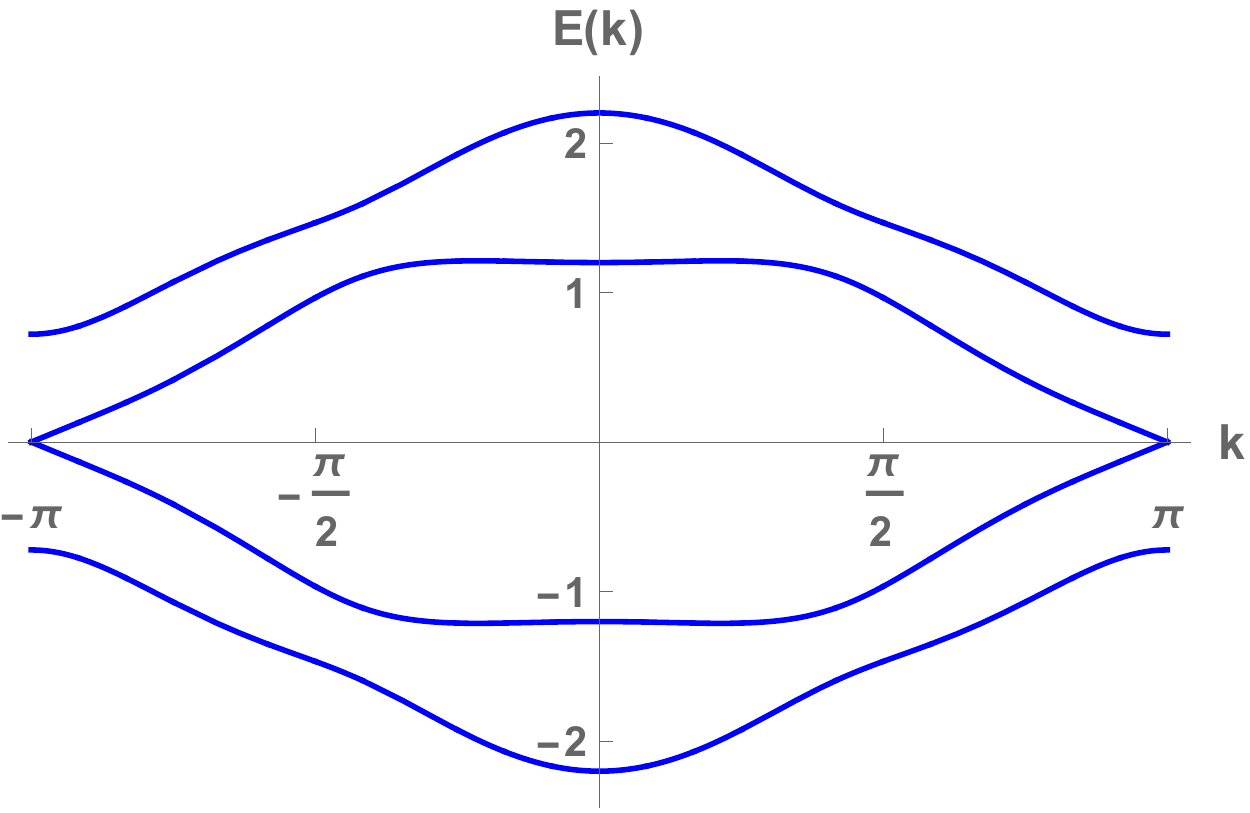}
\caption{Plot of the energies $E_{ik}$ for $i=1,2,3,4$ of
$H_{\alpha}$ within the $4 \times 4$ subspace spanned by the bubble
states $|X_{1(2),k}\rangle$ and $|\Phi_{1(2),k}\rangle$. The top left
(right) panels correspond to $\alpha=0 (0.05)$ and the bottom
left (right) panels to $\alpha=0.1 (0.2)$. Note that the hybridization
spreads over larger region in the Brillouin zone with increasing
$\alpha$.} \label{figpert}
\end{figure}

On turning on $\alpha$, the band hybridizes due to interaction
between states belonging to different $2 \times 2 $ subspaces. The
hybridization occurs for any finite $\alpha$. However at small
$\alpha$, it is significant only in the momentum width $\delta k = 2
\arccos(\alpha/2)$ around $k= \pm 2\pi/3$. Outside this range, the
effect of hybridization is small. This can be clearly seen from top
right and bottom panels of Fig.\ \ref{figpert}. Thus for small
$\alpha \ll 1$, states in the $2 \times 2$ subspace spanned by
$|X_{1(2),k}\rangle$ do not interact with those in the subspace
spanned by $|\phi_{1(2),k}\rangle$ for most momenta within the first
Brillouin zone. We shall use the significance of this fact while
discussing quench dynamics induced by $H_{\alpha}$ in Sec.~\ref{sec:quenches}.

\section{Quench dynamics from some simple initial states}
\label{sec:quenches}
\begin{figure}
 \includegraphics[width=0.96\hsize]{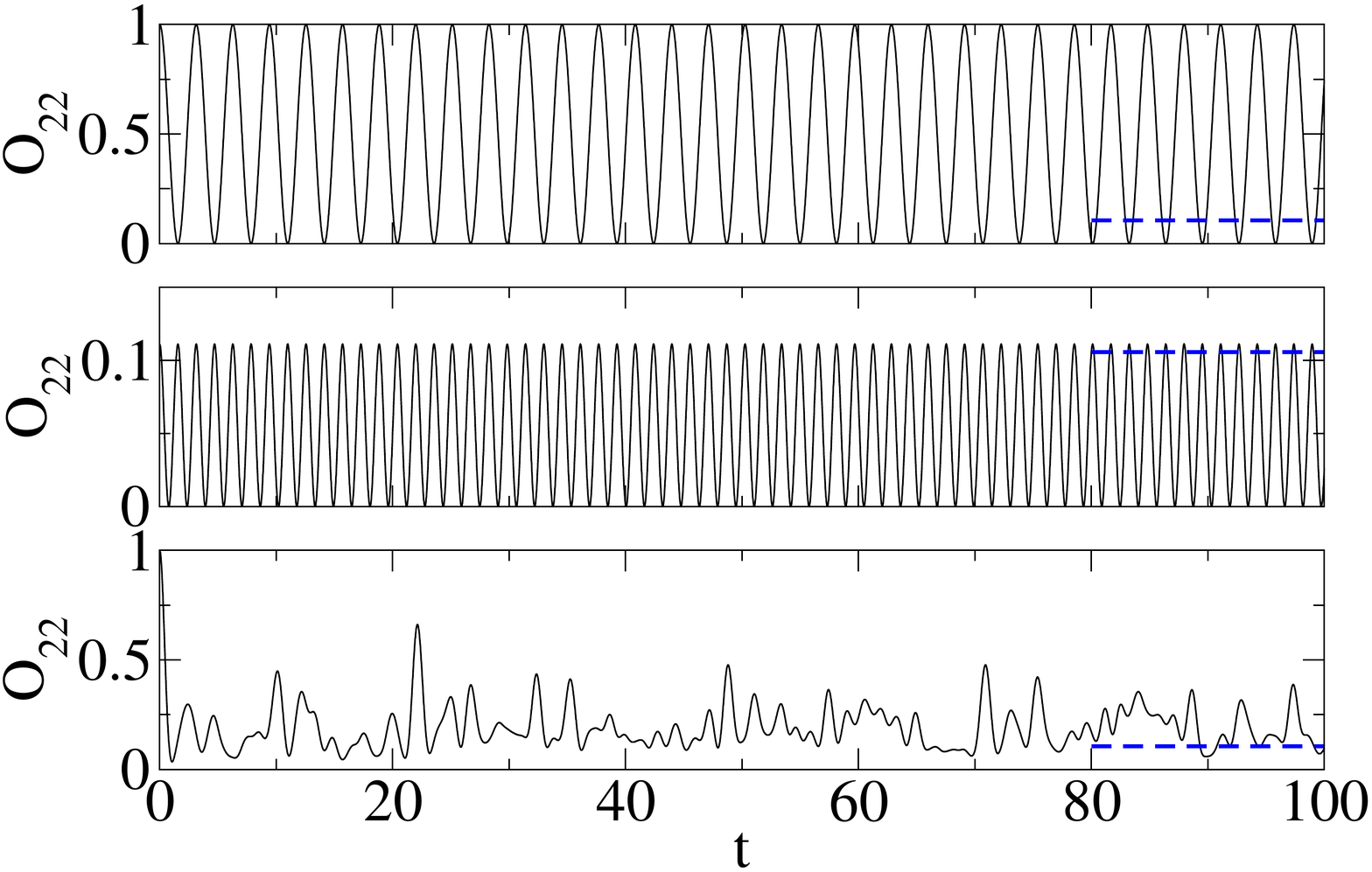}
 \caption{The evolution of a local correlation function as a function of
   time ($t$) for a system of $L=18$ for three different initial states (a
   bubble Fock state with a single $X_{1,j_0}$ at $j_0=3$ in the top panel,
   $|\Phi_{1,k=0}\rangle$ (Eq.~\ref{fragmentink}) in the middle panel and
   $|\mathbb{Z}_2\rangle$ state in the bottom panel)
   quenched
 under $H_3$.}
\label{fig5}
\end{figure}

\begin{figure}
  \includegraphics[width=0.96\hsize]{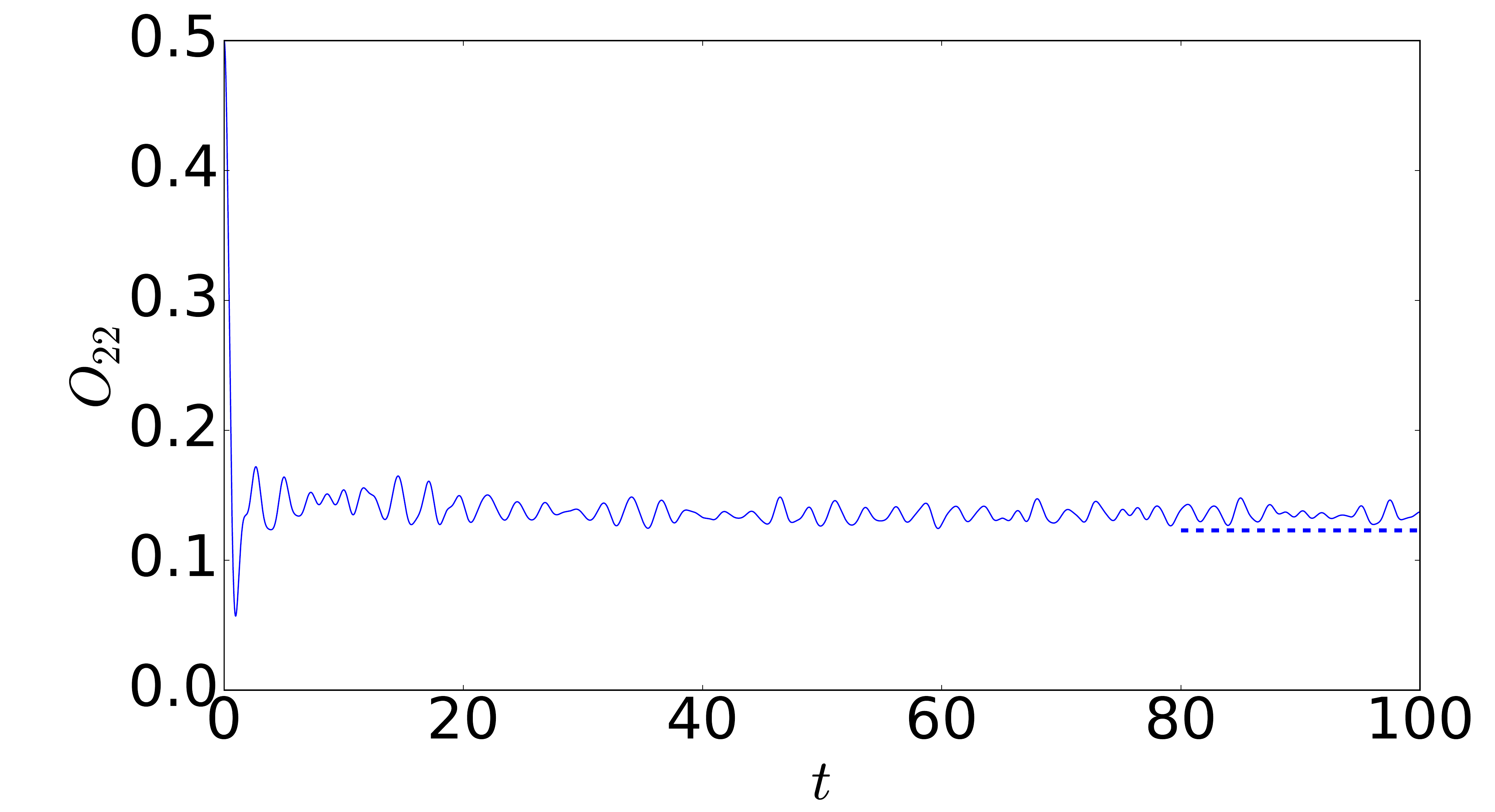}\\
  \includegraphics[width=0.96\hsize]{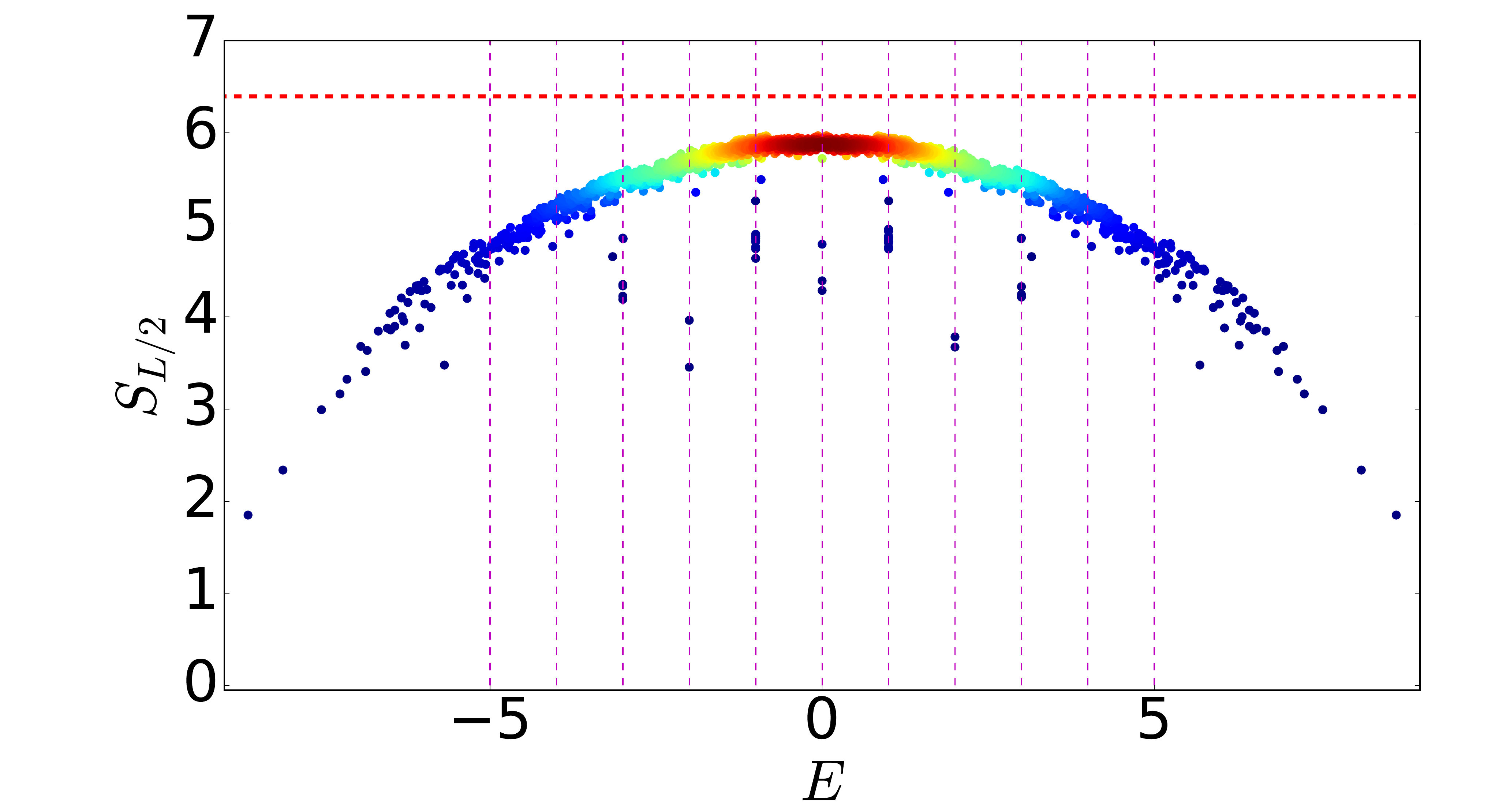}
  \caption{(Top panel) The evolution of a local correlation function as a
    function of time ($t$) for a system of $L=28$ for the initial state
    $(|\mathbb{Z}_2\rangle+|\bar{\mathbb{Z}}_2\rangle)/\sqrt{2}$. (Bottom panel)
    $S_{L/2}$ shown for all the eigenstates in the primary fragment with
    global quantum numbers $k=0, I=+1$ that contains the state
    $(|\mathbb{Z}_2\rangle+|\bar{\mathbb{Z}}_2\rangle)/\sqrt{2}$. The vertical dotted lines are
  given at integer values of $E$ in the bottom panel
  while the horizontal dotted line indicates the average entanglement entropy,
  $S_{\mathrm{Page}}$, of random pure states.}
\label{fig6}
\end{figure}

A dynamical consequence of Hilbert space fragmentation is the lack
of thermalization starting from a class of simple initial states
while other initial states thermalize (since the model is
non-integrable). All the bubble-type Fock states defined in
Sec.~\ref{sec:bubblestates} provide examples of initial states that
do not thermalize under a unitary evolution with $H_3$ and instead
show perfectly coherent oscillations (except the inert Fock state
since its an eigenstate of $H_3$). This phenomenon is easiest to see
for a classical Fock state with a single $X_{1}/X_{2}$ unit (see
Eq.~\ref{onebubbleFock}) since such an initial state only has an
overlap with two bubble-type eigenstates of $H_3$. The
wavefunction at time $t$ starting from a state constituting a single
$X_{1,j_0}$ bubble (denoted by $|X_{1,j_0}\rangle$ for clarity) is given
by
\begin{eqnarray}
|\psi_{1,j_0}(t)\rangle =  \left( \cos t |X_{1,j_0}\rangle + \sin t
|X_{2,j_0}\rangle \right) \label{wavdyn}
\end{eqnarray}
where here and in the rest of this section, we have set $\hbar=1$.
Thus local quantities like $\langle \sigma_j^z \rangle$ (where $j$
coincides with any of the three sites ($j_0,j_0\pm 1$) {within the
bubble) show oscillations. For example, it is easy to see that when
$j_0$ denotes the center site of the bubble
\begin{eqnarray}
\langle \psi_{1j_0}(t)|\sigma_{j_0}^z |\psi_{1j_0}(t)\rangle &=&  -
\cos 2t
\nonumber\\
\langle \psi_{1 j_0}(t)|\sigma_{j_0\pm 1}^z |\psi_{1 j_0)}(t)\rangle
&=& \cos 2t \label{opdyn}
\end{eqnarray}
leading to an oscillation time period $T=\pi$. We note that a
similar oscillation will also be seen for the global magnetization
$M= \langle \sum_j \sigma_j^z \rangle$, albeit about an extensive
constant expectation value, given by
\begin{eqnarray}
M(t) &=& - \left( (L-3) - \cos 2t \right). \label{magdyn}
\end{eqnarray}
where we have taken a chain of length $L$. In contrast, the
staggered magnetization $M_s(t) = \langle \sum_j (-1)^j \sigma_j^z
\rangle$  yields
\begin{eqnarray}
M_s(t) &=& (-1)^{j_0+1} [3 \cos 2t]. \label{staggered}
\end{eqnarray}
Thus $M_s$ exhibits $\pi$ periodic oscillation about zero. In
addition, we shall also be computing the density-density correlation
of the Rydberg atoms between two second-nearest neighbors. These can
be written in terms of the spin operator as
\begin{eqnarray}
O_{j_1 2} = \langle n_{j_1} n_{j_1+2} \rangle = \frac{1}{4} \langle
(1+\sigma_{j_1}^z)(1+\sigma_{j_1+2}^z) \rangle \label{desnop}
\end{eqnarray}
For the bubble states, this correlator also leads to $\pi$ periodic
oscillation
\begin{eqnarray}
O_{j_1 2}(t) &=& \delta_{j_1 j_0-1} \cos^2(t) \label{denosc}
\end{eqnarray}
This is shown numerically for a system of size $L=18$ in
Fig.~\ref{fig5} (top panel) where we have plotted $O_{22}$ with
$j_0=3$. The oscillation time period equals $T=\pi$ for more
complicated bubble Fock states as well which follows from the
``non-interacting'' nature of $H_{3,\mathrm{bubble}}$ in
Eq.~\ref{bubbleHfree}.
Another interpretation of the persistent oscillations
  is that there is an {\it{emergent}} dynamical symmetry that arises
  in the bubble sector since
  \begin{eqnarray}
    [P_{\mathrm{bubble}}H_3P_{\mathrm{bubble}}, \frac{\tau^y_{j_b}+i\tau^z_{j_b}}{2}] =
    \omega\left(\frac{\tau^y_{j_b}+i\tau^z_{j_b}}{2}\right)
    \label{dynsymm}
  \end{eqnarray}
  where $P_{\mathrm{bubble}}$ is a projection operator to the
  bubble Fock
    space and $\omega=2$ given the form of $H_{3,\mathrm{bubble}}$ in
  Eq.~\ref{bubbleHfree}. Thus, for a bubble Fock state, any local operator
  with a finite overlap with the any of the
  $\left(\frac{\tau^y_{j_b}+i\tau^z_{j_b}}{2}\right)$ operators
  will show persistent oscillations~\cite{Medenjak2020}
  with a time period of $T=2\pi/\omega=\pi$.

On the other hand, if one starts with an initial state like
$|\Phi_{1,k=0}\rangle$ (Eq.~\ref{fragmentink}) which is a
zero-momentum version of a non-bubble Fock state, we see that the
corresponding oscillation time scale changes to $T=\pi/2$ because
$\Delta E=4$ in this case (Fig.~\ref{fig5} (middle panel}). Finally,
there are simple initial states that are part of large non-bubble
fragments which seem to thermalize close to the ETH-predicted steady
state value. For example, the $|\mathbb{Z}_2\rangle$ state seems to
be one such state. The temporal behavior of local correlators when
the system is initialized in this state for $L=18$ (see
Fig.~\ref{fig5} (bottom panel)) does not show coherent oscillations
and instead gives a much more complicated dynamics.

To understand this better, we instead use the initial state
$(|\mathbb{Z}_2\rangle +|\bar{\mathbb{Z}}_2\rangle)/\sqrt{2}$. Since
this state belongs to the largest primary fragment of $k=0, I=+1$
sector for the system sizes $L=10,16,22,28\cdots$, the usage of
momentum and spatial inversion symmetries allow us to study the
dynamics for $L=28$ (Fig.~\ref{fig6} (top panel)).
We see that in fact, unlike in the PXP model, the state appears to
thermalize close to an infinite temperature ensemble (calculated
using the states inside the primary fragment) when quenched under
$H_3$, consistent with ETH (Fig.~\ref{fig6}, top panel). In
Fig.~\ref{fig6} (bottom panel), we plot $S_{L/2}$ for the
eigenstates from the primary fragment at $k=0$, $I=+1$ that contains
this particular state. A comparison with Fig.~\ref{fig3} (top left
panel) immediately shows that while the entire symmetry sector
contains several anomalous eigenstates, this particular primary
fragment with the same global quantum numbers contains much fewer
anomalous eigenstates. Moreover, these anomalous eigenstates have
little overlap with $(|\mathbb{Z}_2\rangle
+|\bar{\mathbb{Z}}_2\rangle)/\sqrt{2}$ which explains why the
unitary dynamics under $H_3$ thermalizes this initial state.

The $H_\Delta$ interaction (Eq.~\ref{HDelta}) allows for the
possibility of controlling the energy gaps between different
eigenstates in the bubble-type fragments which can only be integers
when $\Delta=0$. Using any bubble Fock state (except the inert
state) and and following a similar analysis as shown earlier using
$H_{\Delta,\mathrm{bubble}}$ in Eq.~\ref{bubbleHDfree}, we find that
in the presence of $H_\Delta$, the local quantities such as $\langle
\sigma_j^z \rangle$ again shows perfect coherent oscillations with
time period $T = \pi/\sqrt{[1+(9\Delta^2/4)]}$. Thus, $T$ can be
tuned by varying $\Delta$ as shown in Fig.~\ref{fig7} for a
bubble-type Fock state with a single $X_{1,j}$ with $L=18$ centered
around $j_0=3$ where we have plotted $O_{22}(t)$ as a function of
time.

Next, we consider quenching dynamics of bubble Fock states (say
with a single $X_{1,j}$ as shown in Fig.~\ref{fig8} for a chain of
$L=18$) induced by $H_{\alpha}$ (Eq.\ \ref{3pxpham}). In this case,
we find that the dynamics strongly depends of the value of $\alpha$.
When $\alpha=0$, the Hilbert space fragmentation leads to perfect
coherent oscillations with an infinite lifetime (Fig.~\ref{fig5}
(a)). However, for any $\alpha \neq 0$, the presence of the PXP term
in the Hamiltonian destroys the fragmentation. When $\alpha =0.01
\ll 1$ (Fig.~\ref{fig8}), we nonetheless see initial oscillations
which decay with a timescale $t_1^* \sim 125$ followed by a revival.
For $\alpha=0.02$ (Fig.~\ref{fig8}), such revivals are evident first
at $t_1^* \sim 60$ and then again at $t_2^* \sim 140$.
For $\alpha=0.1$, a weak revival is also evident after  $t_1^* \sim
12$ (Fig.~\ref{fig8}). However, for $\alpha=0.5$, we see no
oscillations and the state quickly relaxes to the ETH answer
(Fig.~\ref{fig8}).

To explain the revivals, we consider the limit $\alpha \ll 1$. As
seen from Fig.\ \ref{fig5}, in this limit, the states within the $2
\times 2$ subspace do not hybridize for most parts of the Brillouin
zone. Using this fact and a straightforward analysis similar to the
one carried earlier, we find
\begin{eqnarray}
|\psi_k(t)\rangle \simeq  \cos \epsilon_k t + \sin \epsilon_k t
\end{eqnarray}
provided we start from an initial state $|\psi_k(t=0)\rangle=
|X_{1,k}\rangle$ (where $k$ denotes momentum). Thus the time
evolution of $O_{(j_0-1) 2}(t)$ starting from initial state given by
a single bubble localized at $j=j_0$ is given by
\begin{eqnarray}
O_{(j_0-1)2}(t) & \simeq & \frac{1}{2L}\sum_{\vec k} [1+ \cos (2
\epsilon_k t)] \nonumber\\
&=& \frac{1}{2} [ 1+ \cos(2 t) J_0 (2\alpha t) ] \label{dyn2by2}
\end{eqnarray}
where we have used $\epsilon_k=1 + 2 \alpha \cos k$ and $J_0(x)$
denotes zeroth order Bessel function. Thus for $t_{n}^{\ast} =
x_n/(2 \alpha)$, where $x_n$ denotes the position of $n^{\rm th}$
zero of $J_0(x)$, one expects the oscillations to vanish followed by
a revival for $t>t_n^{\ast}$.

We note that for small $\alpha$, $t_{n}^{\ast} = t^{\ast}$ found
from exact numerics; for example, for $\alpha=0.01$ (first panel of
Fig.\ \ref{fig8}), $t^{\ast}=t_1^{\ast} \simeq 120$ (since
$x_1=2.4$) while for $\alpha=0.02$ (second panel of Fig.\
\ref{fig8}), $t_1^{\ast} \simeq 60$ and $t_2^{\ast} \simeq 139$
(since $x_2 \simeq 5.54$) matches the positions where oscillation
vanishes. For a larger $\alpha=0.1$ (third panel of Fig.\
\ref{fig8}), while $t_1^{\ast} \simeq 12$ matches well with the
numerics, the next revival at $t_2^{\ast} \simeq 28$ is harder to
see because of the decay of the oscillations. For still larger
$\alpha=0.5$ (fourth panel of Fig.\ \ref{fig8}), the local
observable rapidly attains the value. Moreover for $t \ll
\alpha^{-1}$, the oscillations have $\pi$ periodicity as expected
from Eq.\ \ref{dyn2by2}.

While these features are reproduced by this calculation (revival and
initial period of oscillations), there are several aspects of the
dynamics which are not captured by this rather simplistic approach.
For example, we find that $O_{(j_0-1) 2}(t)$ almost touches zero at
several times during the oscillation which can not be explained by
Eq.\ \ref{dyn2by2} even when we take into account the possibility of
its modification by a simple exponential decay factor which may stem
from the loss of weight of the initial state from the $4 \times 4$
subspace. Moreover, Eq.\ \ref{dyn2by2} predicts a value $O_{(j_0-1)
2}(t_n^{\ast}) =1/2$ for any $t_n^{\ast}$; instead exact numerics
yields a value close to zero. We do not have an analytic explanation
of these features of $O_{(j_0-1) 2}(t)$.

\begin{figure}
 \centering
 \includegraphics[width=0.96\hsize]{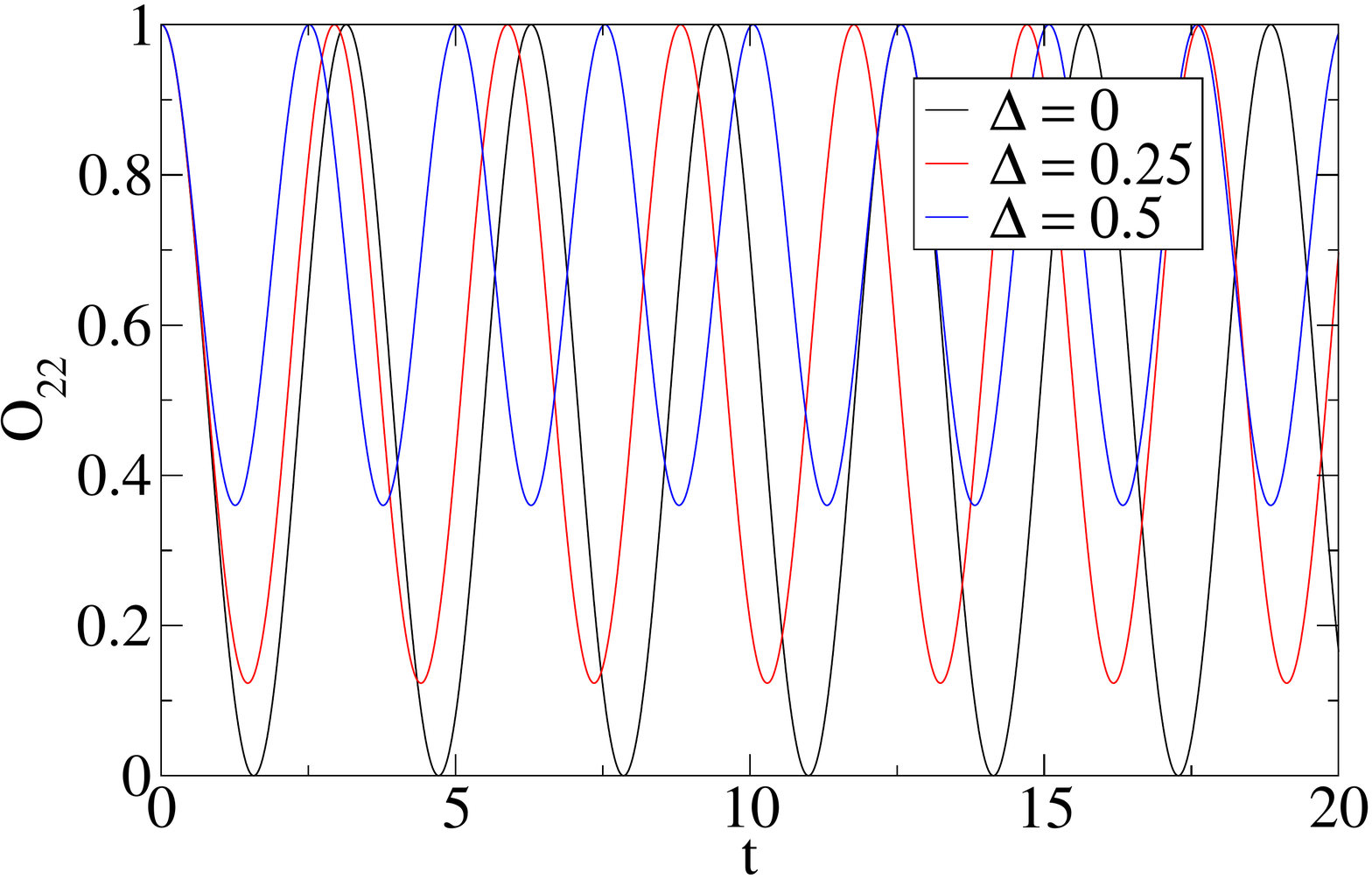}\\
 \caption{Unitary dynamics starting from an initial bubble Fock state
   with a single $X_{1,j_0}$ with $j_0=3$ quenched under $H_\Delta$ for three different values
 of $\Delta$ for a chain of length $L=18$.}
 \label{fig7}
 \end{figure}

 \begin{figure}[!tbp]
 \centering
 \includegraphics[width=0.96\hsize]{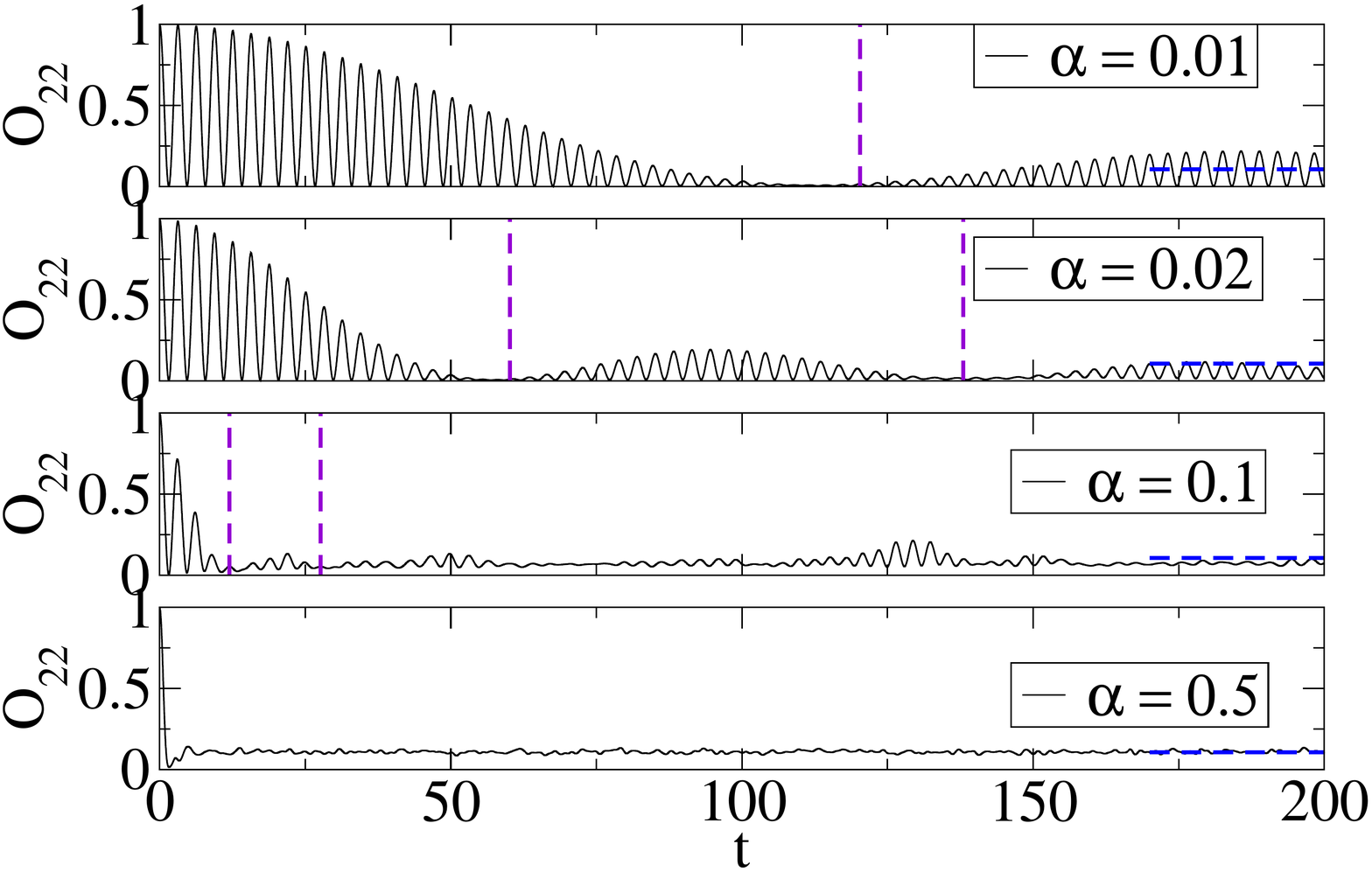}
 \caption{Unitary dynamics starting from an initial bubble Fock state
with a single $X_{1,j}$ quenched under $H_3+\alpha H_{\mathrm{PXP}}$
for four different values of $\alpha$ for a chain length of $L=18$.
The vertical lines in the top three panels are drawn at $t_n^{\ast}=
x_n/(2\alpha)$ where $x_n$ denote the zeroes of the Bessel function
 $J_0(x)$ as a guide to the eye. }
 \label{fig8}
 \end{figure}

\section{Mapping to a lattice gauge theory}

\label{sec:LGT} In this section, we show how to exactly map the
$H_3$ model (Eq.~\ref{H3def}) to a Hamiltonian formulation of a
$U(1)$ lattice gauge theory coupled to fermionic matter in one
dimension and interpret some of the results of the previous sections
in the language of this gauge theory. Such a mapping was already
implemented in Ref.~\onlinecite{SuracePRX} for the PXP model and we
generalize the same to the $H_3$ model here.

\begin{figure}
  \includegraphics[width=0.96\hsize]{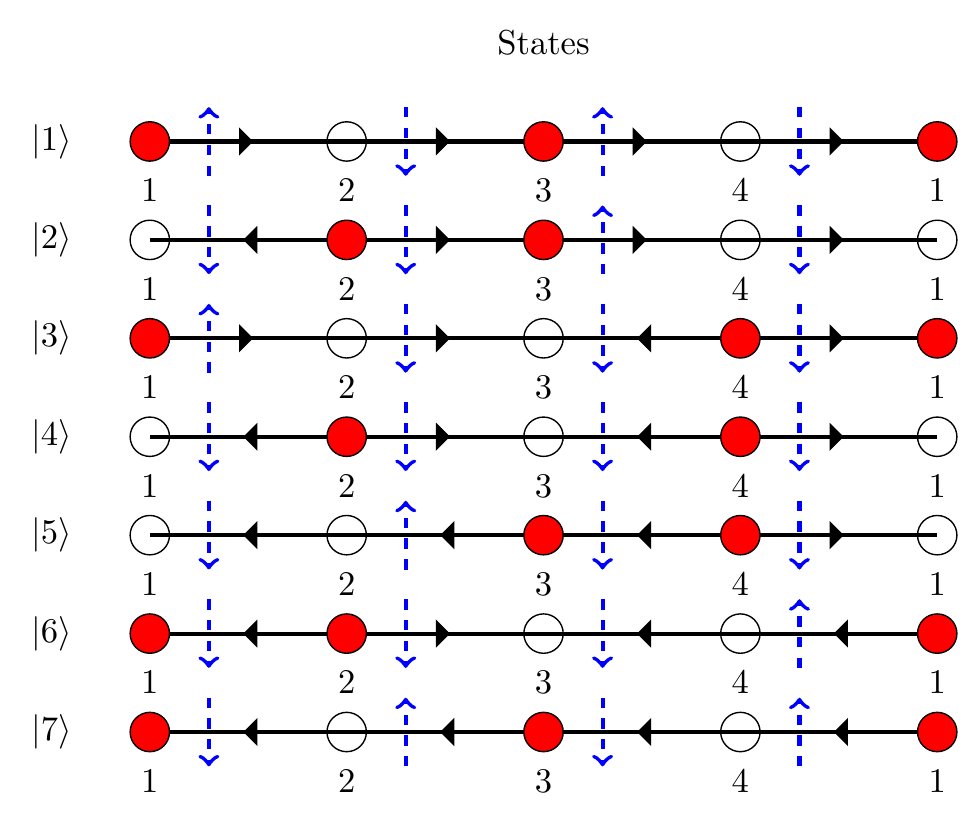}
 \caption{One-to-one correspondence shown between the gauge-invariant states that obey Eq.~\ref{Gauss} and the Fock states in the constrained Hilbert space where no two adjacent $S^z$ can be $\uparrow \uparrow$ simultaneously for a chain length of $L=4$ with periodic boundary conditions. The fermionic matter resides on the sites denoted by circles (where a filled (unfilled) circle represents
   presence (absence) of fermions) while the electric fluxes live on the bonds
   between these circles (denoted by horizontal arrows). The corresponding $S^z=\uparrow,\downarrow$ spins are shown as dotted blue arrows on the dual sites.}
\label{fig9}
\end{figure}

The mapping between the Fock states in the constrained Hilbert space and the
gauge degrees of freedom is illustrated for a $L=4$ system in Fig.~\ref{fig9}.
The gauge degrees live on the links connecting two neighboring sites
$i$ and $i+1$ of a one-dimensional lattice while the fermionic
matter fields live on the sites.
The original spins can be thought to reside on the sites of the dual lattice.
Both the problems are assumed to have periodic boundary conditions.
The gauge degrees are quantum spin $S=1/2$ operators with the electric flux
$E_{i,i+1}=S^z_{i,i+1}$ with a $U(1)$ quantum link, $U_{i,i+1}=S^+_{i,i+1}$ being a
raising operator for it (see Ref.~\onlinecite{QLM} for the formulation of
such quantum link models and their spin-$S$ representation). The electric
flux on the links are related to the original spins on the corresponding
dual sites by the relation
\begin{eqnarray}
  E_{i,i+1}=(1/2)\eta_i \sigma^z_{d(i,i+1)}
  \label{EtoS}
  \end{eqnarray}
where, according to our sign convention (e.g., see Fig.~\ref{fig9}),
$\eta_i=+1$ ($-1$) if $i$ is an
odd (even) site and $\sigma^z_{d(i,i+1)}=\pm 1$ refers to the original spin
residing on the dual site between the link that connects $i$ and $i+1$.
The fermionic creation operator for the matter field at site $i$ is denoted
by $\psi_i^\dagger$ and the corresponding number operator equals
$n_i=\psi_i^\dagger \psi_i$ (whose eigenvalues equal $0$ or $1$).
The matter field at site $i$ is ``slaved'' to the corresponding electric
fluxes on links with $i$ and is defined by the following Gauss law:
\begin{eqnarray}
  E_{i,i+1}-E_{i-1,i}=\psi_i^\dagger \psi_i + \left (\frac{(-1)^i-1}{2} \right).
  \label{Gauss}
\end{eqnarray}
The Gauss law automatically implements the correct kinematic
constraints on the Hilbert space as can be verified by constructing
all the gauge-invariant states (as shown in Fig.~\ref{fig9}) for
$L=4$. Fig.~\ref{fig9} also shows the original spins on the dual
sites for each of these states, showing the one-to-one
correspondence between gauge-invariant states and the Fock states of
the original spins. Thus, the number of gauge-invariant states on a
lattice with $L$ sites with periodic boundary conditions also equals
$F_{L-1}+F_{L+1}$ where $F_j$ is the $j$th Fibonacci number.
Ref.~\onlinecite{SuracePRX} showed that $H_{\mathrm{PXP}}$
(Eq.~\ref{Hpxpdef} can then be written as
\begin{eqnarray}
  H_{\mathrm{PXP}} = \sum_{i=1}^{L}(\psi_i^\dagger U_{i,i+1}\psi_{i+1}+\mathrm{H.c.})
  \label{lgt1}
\end{eqnarray}
which represents a gauge theory with a minimal coupling between the gauge
and matter fields.

Using the exact mapping between the original spins and the gauge
degrees of freedom, the $H_3$ interaction (Eq.~\ref{H3def}) can also
be written as
\begin{eqnarray}
H_3 &=& \sum_{i=1}^{L} (\psi_i^\dagger \Box_{i,i+3} \psi_{i+3} \Gamma_{i+1}\Gamma_{i+2} + \mathrm{H.c.})\nonumber \\
\Box_{i,i+3} &=& U_{i,i+1}U_{i+1,i+2}U_{i+2,i+3}
  \label{lgt2}
  \end{eqnarray}
where $\Gamma_j=n_j$ for odd $j$ and $\Gamma_j=(1-n_j)$ for even
$j$. That Eq.~\ref{lgt2} defines another $U(1)$ lattice gauge theory
may be verified by checking that $[H_3,G_i]=0$ for all $i$ where
$G_i=E_{i,i+1}-E_{i-1,i}-\psi_i^\dagger \psi_i - \left
(\frac{(-1)^i-1}{2} \right)$ are the generators of the $U(1)$ gauge
symmetry. Furthermore, $E_{i,i+1}-E_{i-1,i}=Q_i$ where $Q_i$ defines
the local charge on site $i$ and can equal either $\pm 1$ or $0$.

It is useful to consider the charge configurations associated to the
allowed Fock states in the Hilbert space. Firstly, the reference (or
the inert state) has an alternating configuration of $Q_i=+1$
($Q_i=-1$) charges on even (odd) $i$. Any other valid charge
configuration can be obtained from this reference charge
configuration by selecting pairs of $+1$ and $-1$ charges on
arbitrary sites and annihilating them to form neutral charges $0$ on
the same sites. The lattice gauge theory defined by $H_3$
(Eq.~\ref{lgt2}) induces the following charge dynamics
\begin{eqnarray}
  H_3 &=&\sum_{i \in \mathrm{even}} (|+1_i0_{i+1}0_{i+2}-1_{i+3}\rangle \langle0_i0_{i+1}0_{i+2}0_{i+3}| \nonumber \\
  &+&\mathrm{H.c.}) \nonumber \\
  &+& \sum_{i \in \mathrm{odd}} (|-1_i0_{i+1}0_{i+2}+1_{i+3}\rangle \langle0_i0_{i+1}0_{i+2}0_{i+3}| \nonumber \\
  &+&\mathrm{H.c.}).
  \label{chargedynH3}
\end{eqnarray}

Unlike earlier models~\cite{PaiPN2019, KhemaniN2019, Khemani2020, SalaRVKP2020,
  MoudgalyaPNRB2019} of Hilbert space fragmentation where both total
charge and
dipole-moment were simultaneously conserved, this $U(1)$ gauge
theory only conserves the total charge but not the dipole moment.
This can be easily seen by
taking a charge-neutral configuration with $Q_i=0$ on all sites and then placing a $Q_{2j}=+1$ and $Q_{2j+1}=-1$ on two adjacent sites to create a dipole of
length $l=1$. The charge dynamics induced by $H_3$ (Eq.~\ref{chargedynH3})
then creates longer dipoles leading to a non conservation of the net dipole
moment, as we illustrate below:
\begin{widetext}
  \begin{eqnarray}
    \cdots00\boxed{+1}\boxed{-1}00000000 \rightarrow \cdots00\boxed{+1}\boxed{-1}00\boxed{+1}00\boxed{-1}00\cdots \rightarrow \cdots00\boxed{+1}000000\boxed{-1}00\cdots.
    \label{dipolenon}
    \end{eqnarray}
\end{widetext}

The bubble Fock states defined in Sec.~\ref{sec:bubblestates} have a simple
interpretation in this language. Let us start with the reference configuration
and select any four consecutive sites and make the charges on these four sites
to be zero. This charge configuration is connected to only one other charge
configuration through the dynamics induced by $H_3$ (Eq.~\ref{chargedynH3}):
\begin{eqnarray}
  \cdots&-1&+1-1\boxed{0_j000}+1-1+1\cdots \nonumber \\
  \cdots&-1&+1-1\boxed{+1_j00-1}+1-1+1\cdots
  \label{onebubblecharge}
\end{eqnarray}
and all the charges outside $\boxed{Q_j00Q_{j+3}}$ are completely frozen. These
are the one bubble Fock states that we defined earlier.
On the other hand, the charge dynamics induced by $H_{\mathrm{PXP}}$ (Eq.~\ref{lgt1}
\begin{eqnarray}
 H_{\mathrm{PXP}} &=&\sum_{i \in \mathrm{even}} (|+1_i-1_{i+1}\rangle \langle0_i0_{i+1}|+\mathrm{H.c.}) \nonumber \\
  &+& \sum_{i \in \mathrm{odd}} (|-1_i+1_{i+1}\rangle \langle0_i0_{i+1}|+\mathrm{H.c.}),
  \label{chargedynHpxp}
\end{eqnarray}
does not keep the charges outside $\boxed{Q_j00Q_{j+3}}$ as frozen since there
are several adjacent pairs of $\pm 1 \mp1$ charges provided by the background
of the reference configuration.

To build two bubble Fock states, we simply need to
satisfy the constraint that these units $\boxed{Q_j00Q_{j+3}}$ and
$\boxed{Q_{\ell}00Q_{\ell+3}}$ when inserted in the reference
configuration should have at least one site separating them. The
minimum case with a single site separating the two units is shown
below:
\begin{eqnarray}
  \cdots&+1&-1\boxed{0_j000}+1\boxed{0_{\ell}000}-1+1\cdots \nonumber \\
  \cdots&+1&-1\boxed{+1_j00-1}+1\boxed{0_{\ell}000}-1+1\cdots \nonumber \\
  \cdots&+1&-1\boxed{0_j000}+1\boxed{-1_{\ell}00+1}-1+1\cdots \nonumber \\
   \cdots&+1&-1\boxed{+1_j00-1}+1\boxed{-1_{\ell}00+1}-1+1\cdots \nonumber \\
  \label{twobubblecharge}
\end{eqnarray}
where the charges outside the two units
$\boxed{Q_{j,\ell}00Q_{j+3,\ell+3}}$ are frozen. To form an $n$
bubble Fock state, we simply need to insert $n$ such
$\boxed{Q_j00Q_{j+3}}$ units, where $Q_j=-Q_{j+3}$ and $Q_j$ can be
either $0$ or $+1$ ($-1$) if $j$ is even (odd), and these units are
separated by at least one site in between them with the rest of the
charges following the charge configuration of the reference
configuration. The bubble eigenstates of $H_3$ are then formed using
$(|\boxed{\pm 100\mp1}\rangle \pm |\boxed{0000} \rangle)/\sqrt{2}$
units and the frozen charges as before.

It is interesting to point out that the gauge-invariant states
corresponding to any of the bubble Fock states show a restricted
mobility of the elementary charged excitations under the dynamics
induced by $H_3$ (Eq.~\ref{chargedynH3}). This is because while all
charges outside the $\boxed{Q_j00Q_{j+3}}$ units are completely
immobile, the charges inside the units can fluctuate between
$\boxed{\pm 1 00 \mp 1}$ and $\boxed{0 00 0}$ showing that only
dipoles of length three (in terms of lattice spacing) are mobile.
Thus, the bubble states in the gauge theory show an emergent
fractonic behavior since the charges have a highly reduced mobility.

The non-bubble eigenstates with dispersing quasiparticles as
discussed in Sec.~\ref{sec:nonbubblestates} also have a simple
interpretation in terms of the charge configurations. For this, we
start with the non-bubble Fock state where a unit of six consecutive
sites have zero charge, $\boxed{0_j00000}$, while the rest of the
sites follow the charge configuration of the reference state. The
action of $H_3$ (Eq.~\ref{chargedynH3}) then leads to an effective
hopping of this six-charge unit either in the forward or backward
direction through intermediate charge units $\boxed{\pm1_j0000\mp1}$
that also move forward or backward in the same manner. This is shown
explicitly in Eq.~\ref{bubblemotiondis} for the forward motion of
these units.
\begin{widetext}
  \begin{eqnarray}
    \cdots &-1&+1-1\boxed{0_j00000}+1-1+1 \cdots \leftrightarrows \cdots -1+1-1\boxed{+1_j00-100}+1-1+1 \cdots \leftrightarrows \nonumber \\
    \cdots &-1&+1-1+1\boxed{0_{j+1}00000}-1+1 \cdots \leftrightarrows \cdots -1+1-1+1\boxed{-1_{j+1}00+100}-1+1 \cdots
    \label{bubblemotiondis}
    \end{eqnarray}
  \end{widetext}
These charge configurations provide the exact analog of the Fock
states in Eq.~\ref{mobileFock} and the rest of the steps can be
traced similarly. Looking at the charge configurations in
  Eq.~\ref{bubblemotiondis}, we see that length three dipoles
  attached to two neutral charges (i.e., $\boxed{\pm 100\mp1}00$ or
  alternatively $00\boxed{\pm 100\mp1}$) are the
  effective mobile units here, again
suggesting a connection to fractons. Whether the restricted mobility
of longer, more complicated charge-neutral units give rise to (some
of) the other anomalous states is an interesting open question.

\section{Conclusions and outlook}
\label{sec:conclusions}
In this work, we have considered a minimal model for Hilbert space
fragmentation on a one-dimensional ring defined in terms of $S=1/2$
quantum spins with three-spin interactions between them along with the
hard constraint that no two adjacent spins can have $S^z=\uparrow$
together. Like the prototypical model PXP model,
the many-body spectrum of this model is also reflection-symmetric around
$E=0$ and admits an exponentially large number (in system size) of exact
mid-spectrum zero modes due to an index theorem. Additionally, the spectrum
has exact degeneracies at other integer eigenvalues (after a suitable
choice of Hamiltonian normalization) for arbitrary chain lengths.

In spite of the
non-integrable nature of this model, a class of its
high-energy eigenstates can be shown to violate the eigenstate thermalization
hypothesis in the thermodynamic limit. This is due to a combination of the
kinematic constraints and the nature of interactions which causes the Hilbert
space to fracture into disconnected fragments
that are closed under the action of the Hamiltonian. These
fragments cannot be distinguished by their global quantum numbers alone and the
largest such fragment, though exponentially large in the system size, still
occupies a vanishingly small fraction of the total Hilbert space in the
thermodynamic limit. This fragmentation allows for closed-form
expressions for many of the high-energy eigenstates in terms of quasiparticles.

The simplest class of such eigenstates, the so-called bubble eigenstates, are
composed of strictly localized quasiparticles and have a natural
representation in real space. These eigenstates emerge from fragments
composed exclusively of bubble Fock states with the smallest such fragment
being $1 \times 1$ and the largest being $2^{n_0} \times 2^{n_0}$ where $n_0 \sim
L/5$ with $L$ being the chain length. All the bubble eigenstates have integer
eigenvalues (including zero)
and follow strict area law for entanglement entropy. A lower bound
on the number of such states demonstrates that their number grows
exponentially with system size for integer energies from zero to
$\pm \mathrm{O}(\sqrt{L})$ when $L \gg 1$.

Furthermore, we have been able to write closed-form expressions for a class of
anomalous high-energy eigenstates that belong to fragments generated from
non-bubble Fock states in this model. These eigenstates are all
expressed in terms of mobile quasiparticles and thus have a natural
representation in momentum space. The simplest of these belong to $2 \times 2$
fragments where the basis states are generated using
certain non-bubble Fock states and their translations. These eigenstates
have eigenvalues that lead to two dispersive bands at high energies that
consist of irrational eigenvalues in general when $L \gg 1$.

We have pointed out a novel secondary
fragmentation mechanism in this model whereby a class of eigenstates of
non-bubble type fragments can be analytically calculated even when the
dimension of the fragment grows rapidly with system size. This is
because certain linear combinations of a fixed number of basis states in
momentum space (whose number does not scale with the system size) leads to
formation of a smaller secondary fragment of a fixed dimension inside the
large primary fragment, with the states inside the secondary fragment being
disconnected from the rest of the Hilbert space under the
action of the Hamiltonian. We show the existence of two anomalous eigenstates
(with energies $E=\pm 1$) in the sector with zero momentum and spatial inversion
symmetry and two flat bands (with energies $E=\pm 1$)
in momentum space via this mechanism.

At this point, it is useful to again stress on
  some of the key differences compared to other
  models of Hilbert space fragmentation in the literature. While previous
  models had two simultaneous $U(1)$ conservation laws (with these being
  ``charge'' and ``dipole'' conservations in most cases), this model does not
  have such simultaneous conservations. Rather, the fragmentation is produced
  due to an interplay of interactions and the constrained nature of the Hilbert
  space; conservation laws emerge for specific sectors as discussed in
  Sec.~\ref{sec:LGT}. While previous models had an
  exponentially large number of ``inert'' states which formed $1 \times 1$
  fragments on their own due to the global charge and dipole conservation
  laws, this model has only one such inert state due to the absence of
  dipole conservation (Sec.~\ref{sec:LGT}).
  Lastly, the secondary fragmentation mechanism for this model
  that leads to some anomalous
  eigenstates even in large primary fragments with most states being typical
has not been pointed out in earlier studies to the best of our knowledge.

We considered the effects of adding two different non-commuting
interactions to the minimal model. Both the perturbed models still
continue to have an $E \rightarrow -E$ symmetry and an exponentially
large number (with system size) of zero modes in their many-body
spectra. The perturbation with a staggered magnetic field term
preserves Hilbert space fragmentation and gives the same number of
zero modes for any non-zero and finite value of the
field
based on exact diagonalization results on
small chains ($L \leq 20$). We were able to analytically understand
a class of these zero modes and their eigenstates.
We also defined a Floquet version with a periodically
  driven staggered magnetic field and showed that the  Floquet unitary
  continues to have a large class of eigenstates with area law entanglement.
  The Floquet version has the added feature that all initial bubble
  Fock states show exact stroboscopic freezing at an infinite number of
  special drive frequencies even in the thermodynamic limit.
Another
perturbation with the PXP term is more complicated since it
immediately destroys the Hilbert space fragmentation of the
unperturbed model. However, an approximate treatment was still
possible for certain initial states when the perturbation can be
considered to be small.

Quench dynamics starting from some simple initial states were also
discussed. While bubble Fock states showed coherent
oscillations in local observables such as $O_{22}$,
the frequency of which could be tuned by adding a suitable staggered
magnetic field, some other states like the N\'eel state rapidly
thermalized (unlike in the PXP model). Quenching with the perturbed
model with a PXP term showed qualitative distinctions between the
small perturbation case and otherwise. In the small
perturbation limit, we found $O_{22}$ to exhibit oscillations with
revivals enclosed within a decaying envelope; the position of the
revivals and the time period of the short-time oscillations was
understood using a simple, analytically tractable, model. However,
other details of the dynamics such as the decay rate could not be
understood analytically. With increasing strength of the PXP term,
the oscillatory behavior is lost and we found a rapid decay to the
ETH predicted thermal value.

We showed how to map this model to an exact $U(1)$ quantum link model in its
$S=1/2$ representation interacting with dynamical fermions. The Gauss law in
the gauge theory automatically enforced the constraints in the Hilbert space.
We interpreted some of the anomalous eigenstates in this language and also
showed that the charged degrees of freedom have reduced mobility for these
states with only certain charge-neutral objects being mobile, a tell-tale
signature of fractons.

Several open issues emerge from our study. Exact diagonalization
results on small chains ($L \leq 20$) already show that while the
bubble eigenstates exhaust the count of integer eigenstates close to
$|E| \sim n_0$ with $n_0 \sim L/5$, there are many more non-bubble
eigenstates that are integer-valued in the vicinity of $E=0$ with
their numbers rapidly increasing with system size. While the bubble
eigenstates follow strict area law scaling of entanglement entropy
and we have shown a few examples of non-bubble integer eigenstates
that are anomalous, we believe that the majority of these non-bubble
integer eigenstates in the neighborhood of $E=0$ (with the extend of
$|E| \sim \mathrm{O}(\sqrt{L})$ to be precise) may have volume law
scaling of entanglement entropy when $L \gg 1$. It will be useful to
address this issue both theoretically and numerically using
techniques similar to Ref.~\onlinecite{Karle}. While we have a
complete understanding of the bubble eigenstates in this minimal
model, it will be useful to see if a deeper understanding can be
obtained for the non-bubble anomalous eigenstates, especially those
generated by secondary fragmentation. After mapping this model to a
$U(1)$ lattice gauge theory, we see that it only has charge
conservation unlike earlier fractonic gauge theories, where some
higher moment is also conserved. However, different charge-neutral
units with reduced mobilities arise when focusing on a class of the
anomalous states. Addressing this emergent fractonic physics in this
gauge theory is left for future works. It seems possible to
generalizing this model to two dimensions in multiple ways and it
will be interesting to see whether the higher-dimensional versions
violate the eigenstate thermalization hypothesis as well.

Lastly, it is likely that the Hilbert space fragmentation mechanism
and anomalous dynamics from certain initial states that we have discussed
may be realized using a Floquet setup. Refs.~\onlinecite{dyn1,dyn5}
showed that a certain periodic drive protocol wherein a
time-periodic magnetic field is applied to the PXP model
results in a Floquet Hamiltonian with both PXP and non-PXP type terms with
the $H_3$ interaction being the leading non-PXP interaction with the
subsequent sub-dominant terms becoming more long-ranged
(e.g., an $H_5$ interaction involving five-spin terms, an $H_7$ interaction
involving seven-spin terms, and so on). Ref.~\onlinecite{dyn5} showed that
tuning the drive frequency alone, it is possible to make the coefficient of the
PXP term go to zero in the Floquet Hamiltonian which leads to freezing of
the reference state with all $S^z=\downarrow$. At these drive frequencies, the
Floquet Hamiltonian is also fragmented since the reference state forms a
$1 \times 1$
fragment of its own. To see coherent oscillations from bubble-type Fock
states, a Floquet protocol is needed where the coefficients of both the PXP
and the $H_5$ interactions can be simultaneously tuned to zero.
Since the longer-ranged terms
like $H_7$ etc annihilate any bubble-type Fock states with distant bubbles,
these coefficients need not be fine-tuned.

\begin{acknowledgments}
  B.M., K.S. and A.S.
  are grateful to Sourav Nandy and Diptiman Sen for related
  collaborations. B.M acknowledge funding from Max Planck Partner Grant
at ICTS and the Department of Atomic
Energy, Government of India, under project no. RTI4001 at ICTS-TIFR.
  The work of A.S.
is partly supported through the Max Planck Partner Group program
between the Indian Association for the Cultivation of Science
(Kolkata) and the Max Planck Institute for the Physics of Complex
Systems (Dresden).

\end{acknowledgments}


\begin{thebibliography}{99}

\bibitem{ETH1} J. M. Deutsch, Phys. Rev. A {\bf 43}, 2046 (1991).

\bibitem{ETH2} M. Srednicki, Phys. Rev. E {\bf 50}, 888 (1994).

\bibitem{ETH3} M. Rigol, V. Dunjko, and M. Olshanii, Nature {\bf 452}, 854 (2008).

\bibitem{ETH4} A.~Polkovnikov, K.~Sengupta, A.~Silva, and M.~Vengalattore, Rev. Mod. Phys. {\bf 83}, 863 (2011).

\bibitem{ETH5} L. D'Alessio, Y. Kafri, A. Polkovnikov, and M. Rigol, Advances in Physics {\bf 65}, 239 (2016).

\bibitem{ETH6} P. Reimann, New Journal of Physics {\bf 17}, 055025 (2015).

\bibitem{ETH7} C.~Gogolin and J.~Eisert, Rep. Prog. Phys. {\bf 79}, 056001 (2016).

\bibitem{VR2016} L. Vidmar and M. Rigol, Journal of Statistical Mechanics: Theory and Experiment {\bf 2016}, 064007 (2016).

\bibitem{PH2010} A.~Pal and D.~A.~Huse, Phys. Rev. B {\bf 82}, 174411 (2010).

\bibitem{NH2015} R. Nandkishore and D. A. Huse, Annu. Rev. Condens. Matter Phys. {\bf 6}, 15 (2015).


\bibitem{ryd_exp} H.~Bernien, S.~Schwartz, A.~Keesling, H.~Levine, A.~Omran, H.~Pichler, S.~Choi, A.~S.~Zibrov, M.~Endres, M.~Greiner, V.~Vuleti\'{c}, and M.~D.~Lukin, Nature {\bf 551}, 579 (2017).

\bibitem{pxp1} C.~J.~Turner, A.~A.~Michailidis, D.~A.~Abanin, M.~Serbyn, and Z.~Papi\'{c}, Nat. Phys. {\bf 14}, 745 (2018).

\bibitem{pxp2} C.~J.~Turner, A.~A.~Michailidis, D.~A.~Abanin, M.~Serbyn, and Z.~Papi\'{c}, Phys. Rev. B {\bf 98}, 155134 (2018).

\bibitem{pxp3} S.~Sachdev, K.~Sengupta, and S.~M.~Girvin, Phys. Rev. B {\bf 66}, 075128 (2002).

  \bibitem{pxp3a} P.~Fendley, K.~Sengupta, and S.~Sachdev, Phys. Rev. B {\bf 69}, 075106 (2004).

\bibitem{pxp4} I.~Lesanovsky and H.~Katsura, Phys. Rev. A {\bf 86}, 041601(R) (2012).

  \bibitem{pxp5} T.~Iadecola, M.~Schecter, and S.~Xu, Phys. Rev. B {\bf 100}, 184312 (2019).

\bibitem{aklt1} N.~Shiraishi, J. Stat. Mech. (2019) 083103.

\bibitem{aklt2} S.~Moudgalya, N.~Regnault, and B.~A.~Bernevig, Phys. Rev. B {\bf 98}, 235156 (2018).

\bibitem{aklt3} S.~Moudgalya, S.~Rachel, B.~A.~Bernevig, and N.~Regnault, Phys. Rev. B {\bf 98}, 235155 (2018).

\bibitem{aklt4} D.~K.~Mark, C.~-J.~Lin, and O.~I.~Motrunich, Phys. Rev. B {\bf 101}, 195131 (2020).

\bibitem{aklt5} S.~Moudgalya, E.~O’~Brien, B.~A.~Bernevig, P.~Fendley, and N.~Regnault, Phys. Rev. B {\bf 102}, 085120 (2020).

\bibitem{qh1} S.~Moudgalya, B.~A.~Bernevig, and N.~Regnault, Phys. Rev. B {\bf 102}, 195150 (2020).

\bibitem{qh2} B.~Nachtergaele, S.~Warzel, and A.~Young, J. Phys. A: Math. Theor. {\bf 54} 01LT01 (2021).

\bibitem{HM1} O.~Vafek, N.~Regnault, and B.~A.~Bernevig, SciPost Phys. {\bf 3}, 043 (2017).

\bibitem{HM2} T.~Iadecola and M.~\v{Z}nidari\v{c}, Phys. Rev. Lett. {\bf 123}, 036403 (2019).

\bibitem{HM3} D.~K.~Mark and O.~I.~Motrunich, Phys. Rev. B {\bf 102}, 075132 (2020).

\bibitem{HM4}  S.~Moudgalya, N.~Regnault, and B.~A.~Bernevig, Phys. Rev. B {\bf 102}, 085140 (2020).

\bibitem{mag1} M.~Schecter and T.~Iadecola, Phys. Rev. Lett. {\bf 123}, 147201 (2019).

\bibitem{dyn1} B.~Mukherjee, S.~Nandy, A.~Sen, D.~Sen, and K.~Sengupta, Phys. Rev. B {\bf 101}, 245107 (2020).

\bibitem{dyn2} S.~Sugiura, T.~Kuwahara, and K.~Saito, Phys. Rev. Research {\bf 3}, L012010 (2021).

\bibitem{dyn3} S.~Pai and M.~Pretko, Phys. Rev. Lett. {\bf 123}, 136401 (2019).

\bibitem{dyn4} B.~Mukherjee, A.~Sen, D.~Sen, and K.~Sengupta, Phys. Rev. B {\bf 102}, 014301 (2020).

\bibitem{dyn5} B.~Mukherjee, A.~Sen, D.~Sen, and K.~Sengupta, Phys. Rev. B {\bf 102}, 075123 (2020).

\bibitem{dyn6} H.~Zhao, J.~Vovrosh, F.~Mintert, and J.~Knolle, Phys. Rev. Lett. {\bf 124}, 160604 (2020).

\bibitem{dyn7} K.~Mizuta, K.~Takasan, and N.~Kawakami, Phys. Rev. Research {\bf 2}, 033284 (2020).



\bibitem{2d_ryd1} A.~A.~Michailidis, C.~J.~Turner, Z.~Papi\'{c}, D.~A.~Abanin, and M.~Serbyn, Phys. Rev. Research {\bf 2}, 022065 (2020).

\bibitem{2d_ryd2} C.~-J.~Lin, V.~Calvera, and T.~H.~Hsieh, Phys. Rev. B {\bf 101}, 220304(R) (2020).

\bibitem{geometric1} K.~Lee, R.~Melendrez, A.~Pal, and H.~J.~Changlani, Phys. Rev. B {\bf 101}, 241111(R) (2020).

\bibitem{geometric2} P.~A.~McClarty, M.~Haque, A.~Sen, and J.~Richter, Phys. Rev. B {\bf 102}, 224303 (2020).

\bibitem{LGTscars} D.~Banerjee and A.~Sen, Phys. Rev. Lett. {\bf 126}, 220601 (2021).

\bibitem{scarsreview} M.~Serbyn, D.~A.~Abanin, Z.~Papi\'{c}, Nat. Phys. {\bf 17}, 675(2021).


\bibitem{Chamon2005} C.~Chamon, Phys. Rev. Lett. {\bf 94}, 040402 (2005).

\bibitem{Haah2011} J.~Haah, Phys. Rev. A {\bf 83}, 042330 (2011).

\bibitem{VijayHF2016} S.~Vijay, J.~Haah, and L.~Fu, Phys. Rev. B {\bf 94}, 235157 (2016).

\bibitem{Pretko2017} M.~Pretko, Phys. Rev. B {\bf 95}, 115139 (2017).

 \bibitem{PaiPN2019} S.~Pai, M.~Pretko, and R.~M.~Nandkishore, Phys. Rev. X {\bf 9}, 021003 (2019).

 \bibitem{KhemaniN2019} V.~Khemani, M.~Hermele and R.~Nandkishore, Phys. Rev. B {\bf 101}, 174204 (2020).

   \bibitem{Khemani2020} V.~Khemani, M.~Hermele, and R.~Nandkishore, Phys. Rev. B {\bf 101}, 174204 (2020).

 \bibitem{SalaRVKP2020} P.~Sala, T.~Rakovszky, R.~Verresen, M.~Knap, and F.~Pollmann, Phys. Rev. X {\bf 10}, 011047 (2020).

 \bibitem{MoudgalyaPNRB2019} S.~Moudgalya, A.~Prem, R.~Nandkishore, N.~Regnault, and B.~A.~Bernevig, arXiv:1910.14048.


 \bibitem{YangLGI2020} Z-C.~Yang, F.~Liu, A.~V.~Gorshkov, and T.~Iadecola, Phys. Rev. Lett. {\bf 124}, 207602 (2020).

 \bibitem{Tomasi2019} G.~De Tomasi, D.~Hetterich, P.~Sala, and F.~Pollmann, Phys. Rev. B {\bf 100}, 214313 (2019).
   
   \bibitem{LanglettX2021} C.~M.~Langlett and S.~Xu, Phys. Rev. B {\bf 103}, L220304 (2021).

 \bibitem{MukherjeeCL2020} B.~Mukherjee, Z.~Cai, and W. V.~Liu, Phys. Rev. Research {\bf 3}, 033201 (2021). 

   \bibitem{HahnCL2020} D.~Hahn, P.~A.~McClarty, and D.~J.~Luitz, arXiv: 2104.00692.

   \bibitem{LeePC2020} K.~Lee, A.~Pal, and H.~J.~Changlani, Phys. Rev. B {\bf 103}, 235133 (2021).

     \bibitem{expfragmentation} S.~Scherg, T.~Kohlert, P.~Sala, F.~Pollmann, Bharath H.~M., I.~Bloch, and M.~Aidelsburger, Nat. Communications {\bf 12}, 4490 (2021).

     \bibitem{indexth} M.~Schecter and T.~Iadecola, Phys. Rev. B {\bf 98}, 035139 (2018).

     \bibitem{FETH1} A.~Lazarides, A.~Das, and R.~Moessner, Phys. Rev. E {\bf 90}, 012110 (2014).

     \bibitem{FETH2} P.~Ponte, A.~Chandran, Z.~Papi\'{c}, and D.~A.~Abanin, Ann. Phys. {\bf 353}, 196 (2015).

     \bibitem{FETH3} L.~D'.~Alessio and M.~Rigol, Phys. Rev. X {\bf 4}, 041048 (2014).


     \bibitem{fpt1} A.~Soori and D.~Sen, Phys. Rev. B {\bf 82}, 115432
(2010).
     \bibitem{fpt2} T.~Bilitewski and N.~Cooper, Phys. Rev A {\bf 91}, 063611
(2015).
     \bibitem{fpt3} A.~Sen, D.~Sen, and K.~Sengupta, J. Phys. Cond. Mat. {\bf 33} 443003 (2021).

\bibitem{Page} D.~N.~Page, Phys. Rev. Lett. {\bf 71}, 1291 (1993).

\bibitem{adas1} A. Das, Phys. Rev. B {\bf 82}, 172402 (2010)

\bibitem{dpekker1} S. Mondal, D. Pekker, and K. Sengupta, Europhys.
Lett. {\bf 100}, 60007 (2012).

\bibitem{Medenjak2020} M.~Medenjak, B.~Bu\u{c}a, and D.~Jaksch, Phys. Rev. B {\bf 102}, 041117 (R) (2020).

       \bibitem{SuracePRX} F.~M.~Surace, P.~P.~Mazza, G.~Giudici, A.~Lerose, A.~Gambassi, and M.~Dalmonte, Phys. Rev. X {\bf 10}, 021041 (2020).

       \bibitem{QLM} S.~Chandrasekharan and U.-J.~Wiese, Nucl. Phys. {\bf B492}, 455 (1997).

         \bibitem{Karle} V.~Karle, M.~Serbyn, and A.~A.~Michailidis, Phys. Rev. Lett. {\bf 127}, 060602 (2021).

\end{thebibliography}
\end{document}